\documentclass[revtex4]{emulateapj}

\usepackage{graphicx}
\usepackage{lscape}
\usepackage{amssymb}
\usepackage{epstopdf}
\usepackage{amsmath}
\usepackage{mathrsfs}
\usepackage{float}
\usepackage{natbib}
\usepackage{longtable}
\usepackage{rotating}
\usepackage{lscape}

\begin{document}

\shorttitle{Abundance of RGBs in NGC~6681}
\shortauthors{O'Malley, Chaboyer}

\title{High Resolution Spectroscopic Abundances of Red Giant Branch Stars in NGC~6681\footnote{Includes data gathered with the 6.5-m Magellan Clay Telescope located at Las Campanas Observatory, Chile.} \footnote{Based on observations made with the Southern African Large Telescope (SALT)}}

\author{Erin M. O'Malley}
\affil{Department of Physics and Astronomy, Dartmouth College, Hanover, NH 03784}
\email{Erin.M.O'Malley.GR@dartmouth.edu}

\author{Alexei Kniazev}
\affil{South African Astronomical Observatory, Cape Town, South Africa}
\affil{Southern African Large Telescope Foundation, Cape Town, South Africa}
\affil{Special Astrophysical Observatory, Russian Academy of Sciences, N. Arkhyz, KCh R, 369167, Russia}

\author{Andrew McWilliam}
\affil{Observatories of the Carnegie Institution for Science, 813 Santa Barbara Street, Pasadena, CA 91101}

\and

\author{Brian Chaboyer}
\affil{Department of Physics and Astronomy, Dartmouth College, Hanover, NH 03784}

\begin{abstract}
We obtain high resolution spectra of nine red giant branch stars in NGC\,6681 and perform the first detailed abundance analysis of stars in this cluster.  We confirm cluster membership for these stars based on consistent radial velocities of $214.5\pm3.7$ km/s and find a mean [Fe/H] = $-1.63\pm0.07$ dex and [$\alpha$/Fe] = $0.42\pm0.11$ dex.
Additionally, we confirm the existence of a Na-O anti-correlation in NGC\,6681 and identify two populations of stars with unique abundance trends.  With the use of \emph{HST} photometry from \citet{Sara2007} and \citet{PiottoCMD} we are able to identify these two populations as discrete sequences in the cluster CMD.  Although we cannot confirm the nature of the polluter stars responsible for the abundance differences in these populations, these results do help put constraints on possible polluter candidates.

\end{abstract}


\section{Introduction}
The historic picture of globular clusters (GCs) as simple stellar populations has changed dramatically over the past several decades, with evidence now suggesting that most, if not all, GCs are hosts to multiple stellar populations.  Although the framework of a simple stellar population has allowed for the determination of GC absolute distances and ages \citep[e.g.][]{Grat1997,Chab1998,Carretta2000,Grundahl2002,Grat2003,OMalley2017} as an independent test of current theories of cosmology and stellar evolution, if we are to truly understand the nature of these objects and their part in the formation and evolution of the Milky Way, we must gain better understanding of not only their global properties, but their variations as well.

There is overwhelming evidence, both from photometric and spectroscopic studies, that GCs host multiple stellar populations.  The first spectroscopic evidence for multiple populations was found in the form of abundance variations in C, N, Na, and O in red giant branch (RGB) stars, leading to the identification of C-N and Na-O anti-correlations \citep{Osborn1971,Cohen1978,PWS1983}.  However, as the initial spectroscopic observations could only be obtained for the brightest and most evolved cluster members, it was unclear as to the cause of these variations and the concept of a single stellar population was not overturned.  Paving the way towards a new GC framework was the ubiquitous spectroscopic evidence of Na-O anti-correlations in GCs from the RGB to the main sequence (MS) \citep{Grat2001,Carretta2009} and photometric evidence of discrete stellar populations.

Recent photometric observations of GCs have taken advantage of precision photometry, specifically in the UV passbands, as these are most sensitive to the absorption features of the light elements known to show abundance variations.  These studies have shown the splitting of not only the RGB, but the sub-giant branch (SGB) and MS in GC color-magnitude diagrams (CMDs), highlighting the chemical discontinuities between the individual groups \citep[e.g.][]{Bedin2004,Piotto2007,Piotto2012,Milone2010,Monelli2013}.  The connection between the spectroscopic and photometric evidence, with Na-rich populations being redder in (U-B) and bluer in (B-I) \citep{Milone2012,Monelli2013}, was essential to not only confirming the existence of multiple stellar populations in GCs, but understanding their properties.

One cluster that lacks such dual confirmation is NGC\,6681.  As part of their surveys of multiple stellar populations in GCs, \citet{Monelli2013}, \citet{PiottoCMD} and \citet{Milone2017} have found convincing photometric evidence that suggests NGC\,6681, like all other GCs, is host to at least two distinct stellar populations.  Neither \citet{Monelli2013} nor \citet{Soto2017} discuss their results for NGC\,6681 in any detail.  There is no detailed abundance analysis to show a Na-O anti-correlation, nor any other potential trends in element abundances.  In this paper, we present the first detailed chemical abundance analysis of nine bright RGB stars in NGC\,6681.  

This paper is organized as follows.  In $\S$~\ref{Observations} we present our spectroscopic observations and data reduction.  In $\S$~\ref{Abund} we discuss our analysis method, derive spectroscopic model atmosphere parameters for each of our stars and perform a complete abundance analysis of 23 elements.  The results of our analysis are presented in $\S$~\ref{Results} where we discuss trends in both light and heavy element abundances, including the Na-O anti-correlation.  Finally, we discuss the significance of our results in $\S$~\ref{Discussion} and summarize in $\S$~\ref{Summary}.

\section{Observations and Reduction}\label{Observations}
We obtained high resolution spectra of nine RGB stars in NGC\,6681 with reasonably high signal-to-noise (S/N$\sim100$) using the Magellan Inamori Kyocera Echelle (MIKE) double spectrograph on the 6.5 meter Magellan Clay II telescope at Las Campanas Observatory and the High Resolution Spectrograph (HRS) on the 11 meter South African Large Telescope (SALT) at the South African Astronomical Observatory.  These stars were identified as likely cluster members with relatively high confidence using \emph{HST} ACS GC Treasury project photometry \citep{Sara2007}.  It was not possible to confirm cluster membership prior to the observations.  The target stars were also chosen to be bright ($V\lesssim15.0$ mag) in order to achieve the high signal-to-noise ratio desired for this work.  Although many more bright RGB stars were available to choose from, our targets also needed to be fairly isolated ($\sim3$" separation between neighbors with $V-2\leq V\leq V+2$\,mag) in oder to reduce contamination by nearby stars.

\begin{figure*}
\centering
\includegraphics[width=1.0\textwidth]{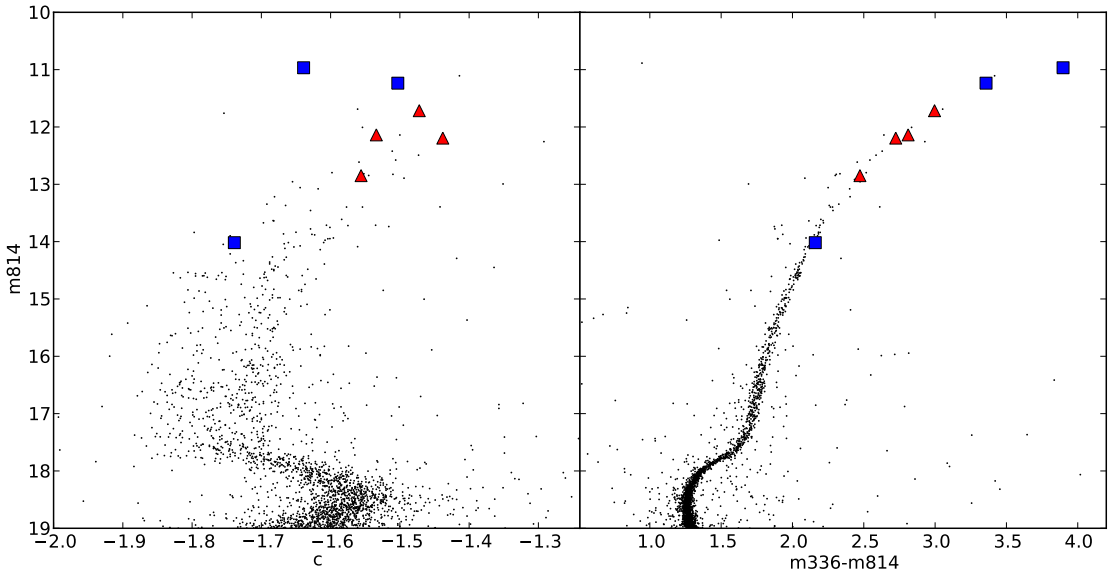}
\caption{Data from \citet{Sara2007}, \citet{PiottoCMD} and \citet{Soto2017} are used to construct a CMD with magnitudes in the \emph{HST} F814W filter.  \emph{Left} - color given as a color index $(m_{275}-m_{336})-(m_{336}-m_{814})$.  A significant spread is seen across the RGB.  Stars in the pristine population (blue squares) fall to the blue side of the RGB while the polluted stars (red triangles) populate the red side. \emph{Right} - a CMD with $m_{336}-m_{814}$ as the x-axis color.} 
\label{fig:Cubi}
\end{figure*}

As can be seen in the color-index diagram and CMD of NGC\,6681 shown in Figure~\ref{fig:Cubi}, the RGB is fairly well-populated and it was our hope that by choosing several targets from small magnitude bins, we would be able to include targets from multiple populations if they existed in the cluster.  With the addition of UV photometry from \citet{PiottoCMD} we are able to construct a color-index diagram to show this in the left panel of Figure~\ref{fig:Cubi}.  Here, the color-index, $c = (m_{275}-m_{336})-(m_{336}-m_{814})$ is used to highlight the separation between the populations.  We find stars in a pristine (blue squares) and polluted (red triangles) population based on Na abundances, which will be discussed further in $\S$~\ref{Results}.

For our MIKE observations we employed a 0.35$\times$5.0\,arcsecond slit, with 2$\times$1 binning, resulting in a measured spectral resolving power of R$\sim$58,000 for the red side and R$\sim$70,000 for the blue side.  Data reduction was performed using the MIKE reduction pipeline \citep{MIKE}.  The MIKE pipeline performs bias subtraction, flat-fields the data and removes background sky signal before extracting the spectral orders from flat traces using optimal extraction algorithms.  The pipeline also performs wavelength calibration based on Th--Ar arc lamp spectra.  The resulting spectra cover a wavelength range of 3315--5000\,\AA\ on the blue side and 4825--9155\,\AA\ on the red side.  We measured the per pixel S/N from the from the rms scatter of the blaze peak near 6730--6750\,\AA~using the IRAF \emph{splot} routine. These high S/N values permit reliable measurement of equivalent widths (EWs) down to $\sim10$\,m\AA.  

SALT HRS is a dual-beam, fiber-fed echelle spectrograph with VPH gratings for cross-dispersers.  Our HRS data cover the wavelength range of 3700--5550\,\AA\ on the blue beam and 5550--8900\,\AA\ on the red beam.  We utilize the spectrograph in the highest resolution mode (R$\sim$67,000--74,000) with a 350\,$\mu$m fiber and the standard instrument readout (single amplifier, 1$\times$1 binning, slow read out) \citep{SALT2014}.  Initial data reduction was performed with the SALT pipeline \citep{Crawford2010}, which includes overscan and gain corrections along with bias subtraction.  The remaining spectroscopic reduction was performed with an HRS pipeline \citep{Kniazev} which makes use of the MIDAS \emph{feros} \citep{SKT1999} and \emph{echelle} \citep{Ballester1992} packages.  The HRS pipeline performs the traditional steps of locating the positions of the blue and red spectral orders from flat frames, identifying and subtracting the 2D background from all frames, extracting the spectrum using optimal extraction algorithms, and finally performing line identification and subsequent wavelength calibration.  Again, the S/N was measured at the blaze peak near 6740\,\AA\ .

For both sets of observations we measure the radial velocities (RVs) of the target stars by cross-correlating our spectra with the Kurucz solar spectrum, using the IRAF \emph{fxcor} routine, and applied heliocentric corrections using the IRAF \emph{rvcorrect} routine.  We estimate the measurement uncertainties to range from 0.9 to 1.8\,km/s.  The derivation of 
RVs serves as check on cluster membership as all cluster members are expected to have similar RVs.  Here we find a mean RV$=214.5\pm3.7$\,km\,s$^{-1}$, which is consistent with the results of \citet{BB1994}, \citet{Rosenberg2000} and \citet{Francis2014}.  

\citet{Soto2017} provide an additional check on the cluster membership for seven of our nine target stars (NGC\,6681\underline{ }7247 and NGC\,6681\underline{ }8125 are excluded in this preliminary data release due to chip orientation) via proper motion estimates.  Targets may be considered cluster members if the displacement between their positions measured by \citet{Sara2007} and \citet{PiottoCMD} are less than 0.35 pixels (in the ACS/WFC pixel frame), $\sim0.02$" based on the ACS Data Handbook estimate of $\sim0.049$" per pixel for the WFC \cite{Lucas2016}.  These measurements cover a roughly eight year timespan and are reliable to 0.005 arcseconds.  The majority of our stars have displacements of less than 0.01", with only one star, NGC\,6681\underline{ } 30380, having a displacement greater than 0.01".  All of the available displacements are below the field cutoff threshold and therefore confirm cluster membership.

A log of the spectroscopic observations along with \emph{HST} F606W and F814W magnitudes converted into V and V-I appears in Table 1.  The stars we attribute to the pristine population based on our Na abundance determinations are listed in bold font and will be discussed in $\S$~\ref{LightTrends}.  We have one target, Star\,12591, in common between the MIKE and HRS data sets which we will use in a comparative analysis.

\begin{table*}
\centering
\setlength{\tabcolsep}{4pt}
{\small
\caption{Magellan and SALT observations of NGC\,6681 RGBs. \label{table:ObsLog}}
\begin{tabular}{l c cl l c c c c r}
\tableline\tableline
ID & RA & Dec & Telescope & Date\hbox{\hskip0.8cm} & Exp. Time & S/N\tablenotemark{a} &V & V-I & RV$_{helio}$\\
 & (deg) & (deg) & & & (s) & (pixel$^{-1}$) & (mag) & (mag) & (km/s)\\
\tableline
 4025 & 280.8066 & -32.3080 & SALT & 12 -- 17 Oct 2016 & 5,805 & 89.8 & 13.51 & 1.35 & 209.9\\
 7247 & 280.7856 & -32.3167 & Magellan & 17 Aug 2015 & 7,500 & 107.4 & 14.50 & 1.19 & 215.4\\
 8125 & 280.7779 & -32.3194 & Magellan & 18 Aug 2015 &10,200 & 98.8 & 15.01 & 1.11 & 219.8\\
10683 & 280.8255 & -32.2974 & SALT & 05 Oct 2015 -- 31 May 2016 & 8,285 & 114.5 & 13.14 & 1.45 & 213.1\\
\textbf{11719} & 280.8193 & -32.3029 & SALT & 05 Jul -- 05 Aug 2015  & 7,440 & 93.3 & 12.80 & 1.59 & 218.8\\
12591 & 280.8160 & -32.2925 & Magellan & 18 Aug 2015 & 3,200 & 125.5 & 13.47 & 1.36 & 211.3\\
&&& SALT & 07 Jun -- 14 Jul 2016 & 9,675 & 105.9 & 13.47 & 1.36 & 211.1\\
12720 & 280.8152 & -32.3020 & Magellan & 17 Aug 2015 & 6,900 & 116.4 & 14.07 & 1.26 &220.0\\
\textbf{30380} & 280.8163 & -32.2836 & SALT & 06 -- 22 Jun 2015  & 7,440 & 76.4 & 12.74 & 1.79 & 210.8\\
\textbf{30847} & 280.8146 & -32.2802 & Magellan & 20 Aug 2015 & 14,000 & 87.6 & 15.11 & 1.13 & 214.8\\
\tableline
\end{tabular}}
\tablenotetext{a}{S/N measured from rms scatter at blaze peak near 6740\AA}
\end{table*}

\section{Abundance Analysis}\label{Abund}
In this study we performed a detailed abundance analysis of nine RGB stars in the GC NGC\,6681.  We adopt the standard spectroscopic approach of determining stellar atmospheric parameters from the equivalent widths (EWs) of Fe\,I and Fe\,II lines from our high S/N spectra.  We enforce abundance equilibrium (i.e. abundance is independent of both line EW and excitation potential) using Fe\,I lines under the assumption of local thermodynamic equilibrium (LTE) to derive stellar excitation temperatures and microturbulent velocities for the target stars.  We also require ionization equilibrium between Fe\,I and Fe\,II to derive a spectroscopic log\,g.

\subsection{Line Lists}
We manually measured the EWs of spectral lines in our target stars using the IRAF \emph{splot} routine with Gaussian fits to the line profiles.  During this process we made a concerted effort to use only single, unblended lines and place the continuum at the appropriate level in order to achieve the most accurate results.  A line list for 23 elements was compiled from a number of studies \citep{Grat2003,Yong2003,Yong2005,AB2005,Marino2015,Koch2016} and is provided in Table~\ref{table:Lines}.  

\begin{deluxetable*}{r r r r r r r r r r r r r r}
\tabletypesize{\small}
\tablecolumns{14}
\tablecaption{Line list for abundance analysis without HFS lines\label{table:Lines}}
\tablewidth{0pt}
\tablehead{
\colhead{Wavelength} & \colhead{Ion} & \colhead{EP} & \colhead{log(\emph{gf})} & \multicolumn{10}{c}{EW (m\AA)}\\
& & & & 4025 & 7247 & 8125 & 10683 & 11719 & 12591$_\mathrm{M}$ & 12591$_\mathrm{S}$ & 12720 & 30380 & 30847\\
}
\startdata
6300.300 & 8.0 & 0.00 & -9.780 & 32.0 & 19.0 & 14.8 & 45.7 & 76.3 & 38.8 & ... & 25.7 & 87.4 & 30.0\\
6363.780 & 8.0 & 0.02 & -10.250 & 11.7 & 4.2 & 3.3 & 21.5 & 39.6 & 6.1 & 14.9 & 14.1 & 70.1 & 6.1\\	
4982.830    &  11.0    &  2.10   & -0.910 & ... & 23.2 & 20.0 & 34.1 & 19.2 & 47.4 & 33.4 & 34.6 & ... & 25.0\\
6154.230    &  11.0    &  2.10   & -1.560 & 15.0 & 6.9 & 10.1 & 15.8 & 16.8 & 18.1 & 22.3 & 13.5 & 16.0 & 6.6\\
6160.750    &  11.0    &  2.10   & -1.260 & 20.7 & 20.9 &  12.1 & 31.7 & 23.4 & 30.3 & 32.2 & 21.4 & 25.2 & 8.0\\
5711.090 & 12.0 & 4.34 & -1.720 & 89.7 & 73.6 & 66.1 & 117.6 & 119.0 & 100.8 & 114.1 & 88.2 & 157.1 & 69.4\\
6318.710 & 12.0 & 5.11 & -1.970 & 11.4 & 15.5 & 18.2 & 23.5 & 31.3& 24.9 & 19.7 & 16.5 & 17.0 & 16.4\\
6319.240 & 12.0 & 5.11 & -2.200 & 19.9 & 5.0 & 8.0 & 14.9 & 20.1 & 6.7 & 16.8 & 3.5 & 19.0 & 9.2\\
$\vdots$ & $\vdots$ & $\vdots$ & $\vdots$ & $\vdots$ & $\vdots$ & $\vdots$ & $\vdots$ & $\vdots$ & $\vdots$ & $\vdots$ & $\vdots$ & $\vdots$ & $\vdots$  \\  
\enddata
\tablecomments{This table is published in its entirety in the electronic edition of the {\it Astrophysical Journal.} A portion is shown here for guidance regarding its form and content.}
\end{deluxetable*}

\begin{figure*}
\centering
\includegraphics[width=0.7\textwidth]{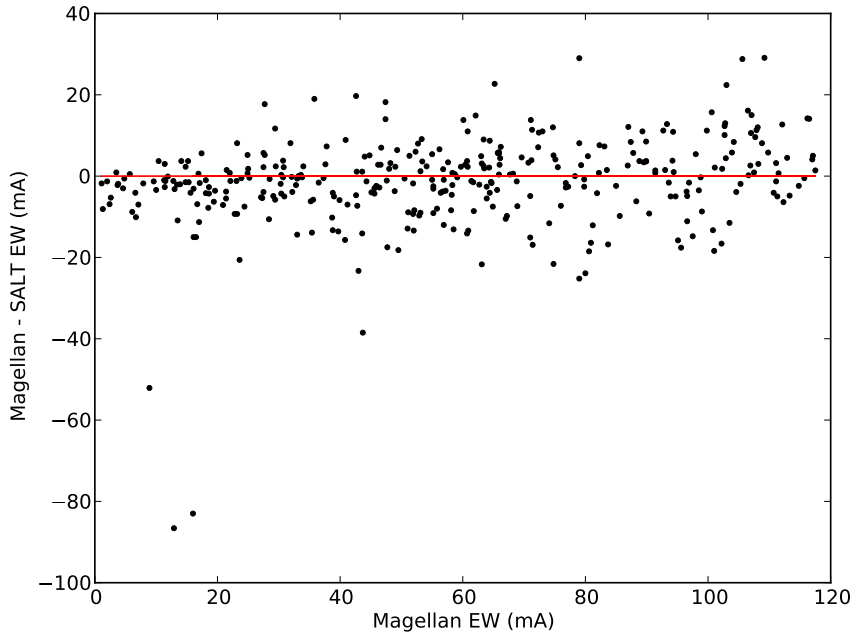}
\caption{Residuals of the EW measurements from the spectra of NGC\,6681\underline{ }12591 taken with the SALT/HRS and Magellan/MIKE spectrographs.  The solid red line depicts zero residual, representing an idealized perfect agreement between the measurements.  Our data follow this line well with no systematic offsets.}
\label{fig:EWcmp}
\end{figure*}

As NGC\,6681\underline{ }12591 has observations from both SALT and Magellan, we show in Figure~\ref{fig:EWcmp} the residuals of the EW measurements from the spectra obtained from each instrument.  We would zero residuals for perfect agreement as depicted by the solid red line.  We find reasonable agreement between the SALT and Magellan EW measurements for NGC\,6681\underline{ }12591 with no systematic offsets.  A further comparison of the abundance measurements for this star is discussed in $\S$~\ref{Results}.

\subsection{Fundamental Stellar Parameters and Atmosphere Models}
We compute the abundance from each line EW using the \emph{abfind} driver of the spectral synthesis program MOOG \citep[2014 version][]{Sneden1973}.  We adopt $\alpha$-enhanced LTE stellar atmospheres generated from the Kurucz grid\footnote{http://kurucz.harvard.edu/grids} \citep{Kurucz} as the spectra of our target stars suggest $\alpha$-enhancement, consistent with their low metallicity \citep{Wall1962,Conti1967,Sneden2004}.

The effective temperature for each of our target stars was determined by requiring there be no trend in Fe\,I abundance as a function of line excitation potential (EP) and as such we call this our excitation temperature, $T_{\mathrm{ex}}$.  The advantage of using excitation temperatures is that they provide stellar temperatures that are independent of reddening.  However, as a first approximation we use the photometric effective temperature, $T_{\mathrm{eff}}$, using the color-temperature relation of \citet{RM05} with the far-infrared reddening values of \citet{DutraBica2000}, which \citet{Chab2017} showed to be a more accurate estimate of cluster reddening than \citet{Harris} based on the color of the RGB.  We find a mean difference between the photometric $T_{\mathrm{eff}}$ and spectroscopic $T_{\mathrm{ex}}$ of $\sim75$\,K.

We use a similar iterative approach to determine the spectroscopic microturbulent velocities, $\xi$, and surface gravities for our stars.  For $\xi$, we make an arbitrary initial guess of 2.0 km/s and iterate until abundance equilibrium is achieved in which the Fe\,I abundance is independent of line EW.  For log\,g, our initial estimate is based on a 12.0 Gyr, $\alpha$-enhanced isochrone from the Dartmouth Stellar Evolution Database \citep[hereafter DSED]{Dott2008} with [Fe/H] = -1.62 and a helium fraction $Y=0.245+1.5Z$.

A comparison of the stellar atmosphere parameters determined photometrically versus spectroscopically is provided in Table~\ref{table:AtmPar}.  We use the spectroscopically determined parameters for the remainder of the analysis.

\begin{table*}
\centering
\caption{Comparison of photometric and spectroscopic stellar atmosphere parameters \label{table:AtmPar}}
\begin{tabular}{l | c c c |c c c c}
\tableline\tableline
 & \multicolumn{3}{c}{Photometric Parameters} & & \multicolumn{3}{c}{Spectroscopic Parameters}\\
\cline{2-4}\cline{6-8}
\tableline
Cluster ID & T$_{\mathrm{eff}}$ (K) & log\,g (dex) & $\xi$ (km/s) & & T$_{\mathrm{ex}}$ (K) & log\,g (dex) & $\xi$ (km/s)\\
\tableline
4025 & 4312 & 1.62 & 2.00 && 4500 & 1.55 & 2.30\\ 
7247 & 4537 & 2.14 & 2.00 && 4610 & 1.20 & 1.90\\
8125 & 4676 & 2.46 & 2.00 && 4650 & 1.40 & 1.80\\
10683 & 4200 & 1.37 & 2.00 && 4250 & 0.80 & 2.15\\
\textbf{11719} & 4075 & 1.06 & 2.00 && 4200 & 0.80 & 2.15\\
12591 Magellan & 4299 & 1.68 & 2.00 && 4250 & 0.65 & 2.10\\
12591 SALT & 4299 & 1.68 & 2.0 && 4300 & 0.75 & 2.10\\
12720 & 4431 & 1.78 & 2.00 && 4510 & 1.20 & 1.90\\
\textbf{30380} & 3951 & 0.72 & 2.00 && 4150 & 0.60 & 3.00\\
\textbf{30847} & 4640 & 2.40 & 2.00 && 4675 & 1.25 & 1.65\\
\tableline
\end{tabular}
\end{table*}

\subsection{non-LTE}
The assumption of local thermodynamic equilibrium (LTE) may not fully capture all of the physical processes occurring within the atmospheres of our target stars.  However, to compute synthetic spectra in non-LTE would be more computationally expensive than the simple LTE approximation.  Therefore, we utilize the INSPECT database\footnote{http://inspect.coolstars19.com/} which takes as inputs the model atmosphere parameters of a star along with the EW of a given line and interpolates over a grid of models to derive a non-LTE correction to our LTE abundances.  We start with corrections to Fe\,I and Fe\,II as we employ those abundances to constrain our model atmosphere parameters.  The calculations of \citet{Berg2012} and \citet{Lind2012} give non-LTE corrections to Fe\,I abundances on the order of 0.06 dex; however, differential relative to the Sun these corrections are reduced and provide non-LTE abundances $\sim$0.03\,dex higher than our LTE results.  For Fe\,II we find similarly small non-LTE corrections of $\sim$0.01 dex, but in the opposite direction of the Fe\,I corrections.

\begin{figure*}
\centering
\includegraphics[width=1.0\textwidth]{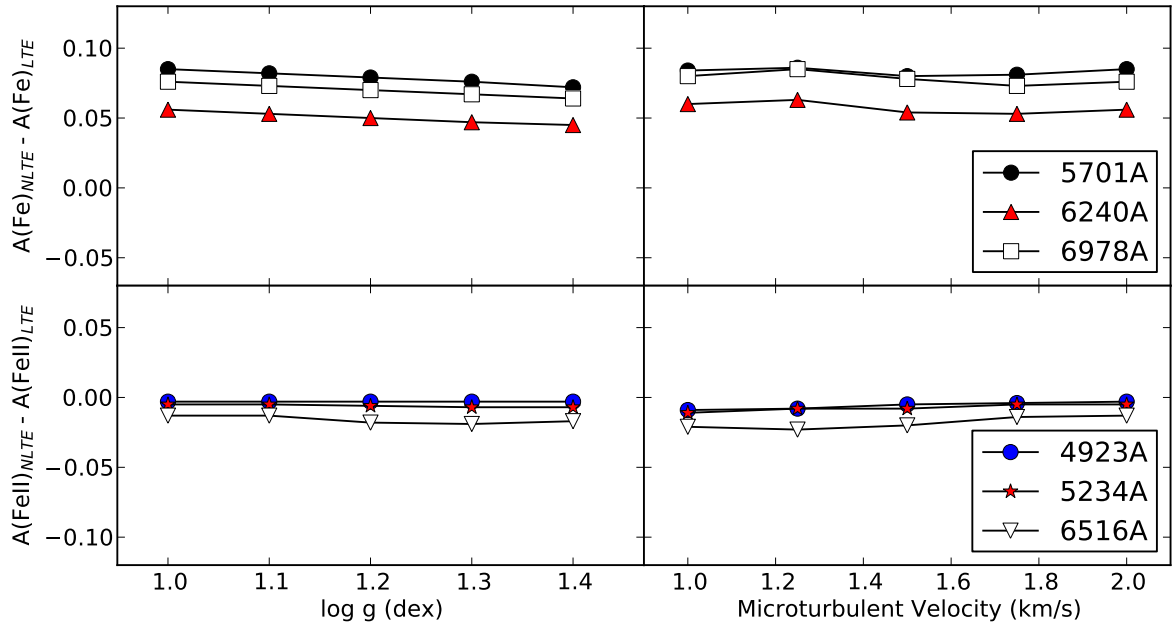}
\caption{non-LTE corrections for Fe\,I (\emph{top}) and Fe\,II (\emph{bottom}) are shown as a function of log\,g (\emph{left}) and microturbulent velocity (\emph{right}).  The corrections are given as $\mathrm{A(Fe)}_\mathrm{non-LTE}-\mathrm{A(Fe)}_\mathrm{LTE}$ and are based on a fixed temperature and EW for NGC\,6681\underline{ }12720 with $\mathrm{[Fe\,I/H]}=-1.61$\,dex and $\mathrm{[Fe\,II/H]}=-1.60$\,dex.}
\label{fig:nltefe}
\end{figure*}

In using the INSPECT tool for the Fe\,I and Fe\,II non-LTE corrections, the range of allowed log\,g ($1.0\leq\mathrm{log}g\leq5.0$\,dex) and microturbulent velocity ($1.0\leq\xi\leq2.0$\,km/s) did not fully cover the range of our target stars.  In Figure~\ref{fig:nltefe} we show the non-LTE corrections for both Fe\,I and Fe\,II as a function of log\,g and $\xi$ for three different lines throughout the stellar spectrum.  We use NGC\,6681\underline{ }12720 as a test case as all of its stellar parameters fall within the range allowed by INSPECT.  As can be seen from Figure~\ref{fig:nltefe}, the size of the non-LTE correction is not highly sensitive to either log\,g or $\xi$ near the range of our target stars; therefore, we are confident that using log\,g = 1.0 dex and $\xi$ = 2.0 km/s for stars with actual parameters outside of the INSPECT range will give approximate corrections that are not substantially different than the actual corrections.

We consult the INSPECT database for non-LTE corrections to Na \citep{Lind2011} and Mg \citep{Osorio2015,OsoBark2015} and, unlike Fe\,I and Fe\,II, we find that the atmosphere parameter range of the model grids fully covers that of our target stars.  For Na, the non-LTE abundance correction is typically -0.08 dex while the absolute non-LTE correction of [Na/Fe] with respect to the Sun is never more than -0.06 dex.  On the other hand, the non-LTE Mg abundance correction in our target stars is small, $\sim0.02$ dex, but increases to 0.06 dex in the absolute [Mg/Fe] non-LTE correction relative to the Sun.  Compared to our typical uncertainties, we find these corrections to be non-negligible and apply them to our final results.

Although many studies have shown that departures from LTE need to be taken into account in deriving Al abundances in stellar spectra \citep{BaumGehren1997,Andrievsky2008,Thyg2014}, currently there is no grid of models that provides corrections over a wide range of stellar parameters.  Therefore, although we note the need for non-LTE corrections, we do not apply them to the Al abundances derived in this study and only estimate approximate corrections.  \citet{Andrievsky2008} give non-LTE abundance corrections for Al on the order of -0.4 dex, near the range of stellar parameters for our target stars ($4700\leq T_{eff}\leq5200$\,K, $\mathrm{log}g = 1.0$\,dex, $\mathrm{[Fe/H]} = -2.0$\,dex).  It is expected that the solar non-LTE correction be small \citep{Smiljanic2016}.  Therefore, we estimate an approximate non-LTE correction of [Al/Fe] relative to the Sun of -0.35 dex.

\subsection{HFS lines}
Although the majority of the abundances in our study were determined using the standard EW analysis, a number of elements (Sc\,II, V, Mn, Co, Cu, Zr, Ba\,II, La\,II and Eu\,II) are affected by hyperfine splitting (HFS) and in some cases isotopic splitting.  The effect of HFS on the measured abundance can depend heavily on the line EW \citep{McWPrest1995}.  Specifically, for strong lines, HFS will desaturate the line, resulting in an over-estimate of the measured abundance.  HFS data were taken from \citet{McWilliam1998}, \citet{Prochaska2000}, \citet{McW2013} and \citet{Thyg2014}.  A portion of the HFS line list is provided in Table~\ref{table:HFS}.

\begin{table*}
\centering
\caption{HFS-elements line list.\label{table:HFS}}
\begin{tabular}{r r r r}
\tableline\tableline
Wavelength & Ion & EP & log(\emph{gf}) \\
\tableline
5657.910 & 21.1 &1.51 &-0.600\\
5669.040 & 21.1 &1.50 &-1.200\\             
5684.190 & 21.1 &1.51 &-1.6021\\
5684.191 & 21.1 &1.51 &-2.0915\\
5684.193 & 21.1 &1.51 &-2.7959\\
5684.204 & 21.1 &1.51 &-1.8962\\
5684.205 & 21.1 &1.51 &-1.9957\\
5684.206 & 21.1 &1.51 &-2.3372\\
5684.215 & 21.1 &1.51 &-2.3372\\
5684.216 & 21.1 &1.51 &-2.0691\\
5684.217 & 21.1 &1.51 &-2.0809\\
6245.621 & 21.1 &1.507 &-1.5820\\
6245.629 & 21.1 &1.507 &-2.3220\\
6245.631 & 21.1 &1.507 &-1.7530\\
6245.636 & 21.1 &1.507 &-3.3220\\
$\vdots$ & $\vdots$ & $\vdots$ & $\vdots$ \\  
\tableline
\end{tabular}
\end{table*}

To treat the HFS properly, we utilized the spectrum synthesis driver \emph{blends} within the MOOG package in order to match the measured EWs to those in a model spectrum that incorporates the blended lines.  It is common to see this HFS analysis performed for lines of Y\,II; however, the effects are typically found to be insignificant and we do not include them here.

\subsection{Abundance uncertainties}
We estimate the systematic uncertainties in our derived abundances due to the stellar atmospheric parameters of our stars by varying the temperature ($\pm50$\,K), surface gravity ($\pm0.2$\,dex), microturbulent velocity ($\pm0.1$\,km/s) and metallicity ($\pm0.10$\,dex) within their respective uncertainties and recomputing the abundances.  We performed this error analysis on each star individually and show the results from NGC\,6681\underline{ }12720 in Table~\ref{table:AbundErr} as it falls roughly in the middle of the parameter space and all elements have been measured.  We add the average abundance offset of each parameter in quadrature to determine the total systematic uncertainty, $\sigma_{par}$. \citet{McWPrest1995} showed the effects of covariance among the atmosphere parameters in abundance determinations and we expect the actual systematic uncertainty to be smaller than those derived in the present work as a result.  

The total uncertainty for each star is then found by adding in the random uncertainty of [X/Fe] and [Fe/H], $\sigma_{\mathrm{rand}} = \sigma/\sqrt(N-1)$, where $N$ is the number of lines measured.  The abundance [X/Fe], along with the uncertainty, is taken with respect to Fe\,I for neutral species and Fe\,II for ionized species and O\,I.  For stellar abundances that are measured from a single line, the random abundance uncertainty is the average of random abundance uncertainties for that species from stars with multiple lines measured.  In the case of Cu and Zr\,II, in which only one line is measured in every star, the random abundance uncertainty is taken to be the random abundance uncertainty of all neutral and singly ionized elements heavier than Fe in the star, respectively.  The total uncertainties are listed along with the mean abundances for each star in Tables~\ref{table:MagAbund} and \ref{table:SALTAbund}, as will be discussed in the following section.

\begin{table*}
\centering
\caption{NGC\,6681\underline{ }12720 abundance changes based on uncertainties on atmosphere parameters. \label{table:AbundErr}}
\begin{tabular}{l c c c c c c c c c r}
\tableline\tableline
Element & \multicolumn{2}{c}{$\Delta T_{\mathrm{ex}}$ (K)} & \multicolumn{2}{c}{$\Delta$log $g$ (dex)} & \multicolumn{2}{c}{$\Delta\xi$ (km s$^{-1}$)} & \multicolumn{2}{c}{$\Delta$[M/H] (dex)} & $\sigma_{\mathrm{par}}$ & $\sigma_\mathrm{rand}$\\
 & +50 & -50 & +0.2 & -0.2 & +0.10 & -0.10 & +0.10 & -0.10 & &\\
\tableline
$\Delta$[O/Fe] & 0.00 & -0.05 & -0.02 & 0.07 & -0.01 & 0.02 & 0.01 & 0.00 & 0.05 & 0.06\\
$\Delta$[Na/Fe] & -0.01 & 0.01 & -0.01 & 0.01 & 0.00 & 0.01 & 0.01 & -0.01 & 0.02 & 0.05\\
$\Delta$[Mg/Fe] & -0.01 & 0.02 & 0.00 & 0.00 & 0.00 & 0.01 & 0.01 & 0.00 & 0.02 & 0.03\\
$\Delta$[Al/Fe] & -0.02 & 0.01 & -0.01 & 0.00 & 0.00 & 0.00 & 0.01 & -0.01 & 0.02 & 0.06\\
$\Delta$[Si/Fe] & -0.04 & 0.01 & -0.01 & 0.05 & -0.01 & -0.03 & -0.01 & -0.03 & 0.05 & 0.03\\
$\Delta$[Ca/Fe] & -0.02 & 0.02 & -0.03 & 0.02 & -0.03 & -0.01 & -0.02 & 0.02 & 0.04 & 0.03\\
$\Delta$[Sc\,II/Fe] & 0.00 & -0.03 & -0.02 & 0.07 & -0.01 & -0.01 & -0.01 & 0.01 & 0.05 & 0.05\\
$\Delta$[Ti\,I/Fe] & 0.02 & -0.02 & -0.01 & 0.01 & 0.00 & 0.01 & 0.00 & 0.00 & 0.02 & 0.03\\
$\Delta$[Ti\,II/Fe] & -0.06 & 0.09 & -0.08 & 0.12 & -0.03 & 0.06 & -0.07 & 0.04 & 0.14 & 0.06\\
$\Delta$[V/Fe] & 0.02 & -0.03 & -0.10 & 0.01 & 0.00 & 0.01 & 0.01 & -0.01 & 0.06 & 0.01\\
$\Delta$[Cr/Fe] & 0.02 & -0.03 & -0.02 & 0.01 & -0.02 & 0.02 & -0.03 & 0.03 & 0.05 & 0.04\\
$\Delta$[Fe\,I/H] & 0.04 & -0.03 & 0.00 & 0.01 & 0.00 & 0.00 & -0.01 & 0.01 & 0.04 & 0.02\\
$\Delta$[Fe\,II/H] & 0.00 & 0.03 & 0.07 & -0.07 & 0.03 & -0.02 & 0.00 & 0.01 & 0.08 & 0.04\\
$\Delta$[Mn/Fe] & 0.00 & -0.02 & -0.02 & 0.01 & -0.01 & 0.01 & -0.01 & 0.01 & 0.02 & 0.08\\
$\Delta$[Co/Fe] & 0.00 & 0.00 & 0.01 & 0.00 & 0.01 & 0.00 & 0.01 & 0.00 & 0.01 & 0.08\\
$\Delta$[Ni/Fe] & 0.00 & 0.01 & 0.01 & 0.00 & 0.01 & 0.01 & 0.01 & 0.01 & 0.02 & 0.03\\
$\Delta$[Cu/Fe] & -0.04 & -0.05 & -0.03 & -0.05 & -0.03 & -0.04 & -0.03 & -0.04 & 0.08 & N/A\\
$\Delta$[Y\,II/Fe] & 0.00 & -0.04 & -0.02 & -0.07 & -0.01 & -0.01 & 0.00 & 0.02 & 0.05 & 0.06\\
$\Delta$[Zr\,I/Fe] & -0.01 & 0.07 & -0.04 & -0.03 & 0.04 & -0.03 & -0.03 & 0.04 & 0.07 & N/A\\
$\Delta$[Zr\,II/Fe] & -0.04 & 0.08 & -0.06 & 0.11 & 0.05 & -0.05 & -0.03 & 0.04 & 0.12 & N/A\\
$\Delta$[Ba\,II/Fe] & 0.03 & 0.00 & 0.08 & -0.07 & 0.05 & -0.02 & -0.06 & 0.09 & 0.10 & 0.06\\ 
$\Delta$[La\,II/Fe] & -0.04 & -0.02 & -0.06 & 0.11 & -0.01 & -0.06 & -0.03 & 0.04 & 0.10 & N/A\\
$\Delta$[Ce\,II/Fe] & 0.01 & -0.04 & -0.02 & 0.00 & 0.00 & -0.01 & 0.01 & 0.01 & 0.03 & 0.04\\
$\Delta$[Nd\,II/Fe] & -0.08 & 0.07 & 0.02 & -0.04 & 0.00 & -0.05 & -0.03 & 0.03 & 0.09 & 0.04\\
$\Delta$[Eu\,II/Fe & 0.00 & 0.03 & 0.11 & -0.10 & 0.05 & -0.02 & 0.01 & 0.02 & 0.09 & N/A\\
\tableline
\end{tabular}
\end{table*}

\section{Results}\label{Results}
Using the EWs measured for each spectral line and Kurucz model atmospheres constructed with the stellar atmosphere parameters given in Table~\ref{table:AtmPar}, we derive the abundances reported in Tables~\ref{table:MagAbund} and \ref{table:SALTAbund} for the Magellan and SALT targets, respectively.  We find an average [Fe/H] = $-1.63\pm0.07$\,dex, in agreement with that of \citet{Carretta2009b} who recalibrate the \citet{Rutledge97} value based on well determined metallicities in other clusters.

\begin{table*}
\centering
\caption{Magellan star abundances. The [X/Fe] ratios employ Fe\,II for ionized species and O\,I. \label{table:MagAbund}}
\begin{tabular}{lrrccccccccccccc}
\tableline\tableline
& \multicolumn{3}{c}{7247} & \multicolumn{3}{c}{8125} & \multicolumn{3}{c}{12591} & \multicolumn{3}{c}{12720} & \multicolumn{3}{c}{30847}\\
Element & Abund & N & $\sigma_{tot}$ & Abund & N & $\sigma_{tot}$ & Abund & N & $\sigma_{tot}$ & Abund & N & $\sigma_{tot}$ & Abund & N & $\sigma_{tot}$\\
\tableline
$$[O/Fe] & 0.27 & 1 & 0.12 & 0.28 & 1 &0.13 & 0.32 & 1 & 0.12 & 0.36 & 2 & 0.09 & 0.55 & 1 &0.11\\
$$[Na/Fe] & 0.24 & 2 & 0.12 & 0.19 & 3 &0.17 & 0.36 & 3 & 0.06 & 0.23 & 3 & 0.05 & 0.02 & 2 & 0.09\\
$$[Mg/Fe] & 0.48 & 2 & 0.14 & 0.58 & 3 & 0.16 & 0.58 & 2 & 0.11 & 0.47 & 2 & 0.04 & 0.51 & 3 & 0.10\\
$$[Al/Fe] & 0.72 & 2 & 0.14 & 0.69 & 2 & 0.15 & 0.58 & 2 & 0.11 & 0.61 & 2 & 0.06 & 0.37 & 1 & 0.12\\
$$[Si/Fe] & 0.41 & 16 &0.06 & 0.42 & 14 & 0.14 & 0.43 & 16 & 0.07 & 0.41 & 14 & 0.06 & 0.38 & 16 & 0.06\\
$$[Ca/Fe] & 0.46 & 17 & 0.08 & 0.44 & 18 & 0.11 & 0.42 & 14 & 0.06 & 0.39 & 14 & 0.05 & 0.44 & 19 & 0.06\\
$$[Sc\,II/Fe] & 0.12 & 5 & 0.08 & 0.04 & 4 & 0.05 & 0.10 & 5 & 0.10 & 0.12 & 4 & 0.07 & 0.00 & 3 & 0.08\\
$$[Ti\,I/Fe] & 0.24 & 24 & 0.06 & 0.18 & 21 & 0.08 & 0.26 & 27 & 0.04 & 0.30 & 30 & 0.04 & 0.21 & 24 & 0.05\\
$$[Ti\,II/Fe] & 0.25 & 6 & 0.14 & 0.16 & 6 & 0.12 & 0.25 & 6 & 0.12 & 0.27 & 6 & 0.16 & 0.28 & 6 & 0.16\\
$$[V/Fe] & 0.13 & 3 &0.10 & 0.03 & 2 & 0.06 & 0.17 & 4 & 0.06 & 0.13 & 4 & 0.06 & -0.04 & 4 & 0.07\\
$$[Cr/Fe] &-0.03 & 4 &0.06 & -0.02 & 4 & 0.07 & -0.15 & 4 & 0.04 & 0.02 & 4 & 0.06 & -0.02 & 4 & 0.06\\
$$[Mn/Fe] & -0.35 & 5 & 0.06 & -0.30 & 5 & 0.07 & -0.20 & 6 & 0.10 & -0.27 & 6 & 0.06 & -0.37 & 3 & 0.04\\
$$[Fe\,I/H]  & -1.66 & 78 & 0.05 & -1.70 & 86 & 0.07 & -1.65 & 66 & 0.07 & -1.61 & 74 & 0.04 & -1.66 & 89 & 0.05\\
$$[Fe\,II/H] & -1.67 & 11 & 0.09 & -1.69 & 10 & 0.06 & -1.64 & 11 & 0.08 & -1.60 & 11 & 0.08 & -1.67 & 11 & 0.07 \\
$$[Co/Fe] & 0.20 & 3 & 0.10 & 0.07 & 1 & 0.08 & 0.13 & 6 & 0.05 & 0.12 & 5 & 0.13  & -0.12 & 5 & 0.08\\
$$[Ni/Fe] & -0.04 & 19 & 0.05 & -0.11 & 17 & 0.07 & -0.04 & 22 & 0.04 & -0.03 & 22 & 0.04 & -0.09 & 16 & 0.04 \\
$$[Cu/Fe] & -0.56 & 1 & 0.10 & -0.79 & 1 & 0.09 & -0.35 & 1 & 0.11 & -0.08 & 1 & 0.10 & -0.06 & 1 & 0.12\\
$$[Y\,II/Fe] & -0.06 & 3 & 0.13 & -0.10 & 6 & 0.12 & 0.05 & 6 & 0.11 & -0.10 & 7 & 0.08 & -0.19 & 7 & 0.10\\
$$[Zr\,I/Fe] & 0.19 & 1 & 0.12 & ... & ... & ... & 0.31 & 2 & 0.07 & 0.44 & 1 & 0.10 & ... & ... & ...\\
$$[Zr\,II/Fe] & 0.27 & 1 & 0.16 & ... & ... & ... & 0.58 & 1 & 0.12 & 0.46 & 1 & 0.13 & 0.25 & 1 & 0.11\\
$$[Ba\,II/Fe] & -0.05 & 3 & 0.12 & -0.11 & 3 & 0.09 & -0.13 & 3 & 0.09 & 0.00 & 3 & 0.11 & 0.02 & 2 & 0.08 \\
$$[La\,II/Fe] & 0.30 & 1 & 0.15 & 0.34 & 2 & 0.11 & 0.24 & 2 & 0.10 & 0.15 & 1 & 0.12 & ... & ... & ...\\
$$[Ce\,II/Fe] & -0.08 & 3 & 0.20 & -0.17 & 1 & 0.12 & -0.02 & 3 & 0.16 & -0.06 & 1 & 0.12 & 0.36 & 3 & 0.32\\
$$[Nd\,II/Fe] &0.08 & 8 & 0.15 & 0.04 & 7 & 0.12 & 0.19 & 7 & 0.12 & 0.15 & 8 & 0.10 & 0.02 & 7 & 0.05\\
$$[Eu\,II/Fe] & 0.12 & 1 & 0.15 & -0.01 & 1 & 0.12 & 0.32 &1 &0.12 & 0.18 & 1 & 0.12 & 0.15 & 1 & 0.11\\
\tableline
\end{tabular}
\end{table*}

\begin{table*}
\centering
\caption{SALT star abudances.  The [X/Fe] ratios employ Fe\,II for ionized species and O\,I. \label{table:SALTAbund}}
\begin{tabular}{lccccccccccccccc}
\tableline\tableline
& \multicolumn{3}{c}{4025} & \multicolumn{3}{c}{10683} & \multicolumn{3}{c}{11719} & \multicolumn{3}{c}{12591} & \multicolumn{3}{c}{30380}\\
Element & Abund & N & $\sigma_{tot}$ & Abund & N & $\sigma_{tot}$ & Abund & N & $\sigma_{tot}$ & Abund & N & $\sigma_{tot}$ & Abund & N & $\sigma_{tot}$\\
\tableline
$$[O/Fe] & 0.48 & 2 & 0.15 & 0.37 & 2 & 0.11 & 0.65 & 2 & 0.12 & 0.20 & 1 & 0.12 & 0.61 & 2 & 0.11\\
$$[Na/Fe] & 0.18 & 2 & 0.15& 0.16 & 3 & 0.12 & 0.01 & 3 & 0.11 & 0.40 & 3 & 0.10 & 0.07 & 2 & 0.08\\
$$[Mg/Fe] & 0.49 & 3 & 0.15 & 0.55 & 3 & 0.07 & 0.53 & 3 & 0.05 & 0.61 & 3 & 0.11 & 0.50 & 2 & 0.15\\
$$[Al/Fe] & 0.59 & 1 & 0.12 & 0.46 & 2 & 0.07 & 0.31 & 3 & 0.15 & 0.60 & 2 & 0.11 & 0.51 & 2 & 0.13\\
$$[Si/Fe] & 0.57 & 14 & 0.09 & 0.43 & 12 & 0.08 & 0.31 & 13 & 0.07 & 0.48 & 16 & 0.09& 0.39 & 10 & 0.08\\
$$[Ca/Fe] & 0.48 & 13 & 0.10 & 0.48 & 7 & 0.08 & 0.43 & 9 & 0.07 & 0.38 & 11 & 0.08 & 0.32 & 5 & 0.13\\
$$[Sc\,II/Fe] &-0.02 & 4 & 0.17 & 0.18 & 4 & 0.08 & 0.27 & 4 & 0.10 & 0.07 & 5 & 0.06 & 0.15 & 4 & 0.06\\
$$[Ti\,I/Fe] & 0.36 & 30 & 0.06 & 0.34 & 26 & 0.05 & 0.54 & 23 & 0.05 & 0.27 & 27 & 0.04 & 0.60 & 20 & 0.06\\
$$[Ti\,II/Fe] & 0.33 & 4 & 0.15 & 0.35 & 5 & 0.13 & 0.51 & 3 & 0.18 & 0.23 & 6 & 0.15 & 0.46 & 4 & 0.15\\
$$[V/Fe] & 0.38 & 3 & 0.06 & 0.08 & 3 & 0.07 & 0.23 & 4 & 0.14 & 0.21 & 3 & 0.06 & 0.04 & 2 & 0.06\\
$$[Cr/Fe] & ... & ... & ... & -0.39 & 1 & 0.07 & ... & ... & ... & -0.23 & 4 & 0.11 & ... & ... & ...\\
$$[Mn/Fe] & -0.08 & 3 & 0.05 & -0.19 & 4 & 0.09 & -0.23 & 4 & 0.10 & -0.15 & 3 & 0.04 & 0.15 & 3 & 0.08\\
$$[Fe\,I/H] & -1.65 & 65 & 0.07 & -1.58 & 56 & 0.09 & -1.55 & 47 & 0.08 & -1.61 & 62 & 0.07 & -1.66 & 41 & 0.08\\
$$[Fe\,II/H] & -1.64 & 10 & 0.09 & -1.58 & 7 & 0.11 & -1.53 & 9 & 0.11 & -1.60 & 10 & 0.11 & -1.69 & 9 & 0.10\\
$$[Co/Fe] & 0.01 & 3 & 0.12 & 0.20 & 7 & 0.08 & 0.35 & 4 & 0.06 & 0.12 & 6 & 0.08 & 0.25 & 4 & 0.04\\
$$[Ni/Fe] & 0.13 & 21 & 0.05 & 0.02 & 23 & 0.05 & -0.02 & 20 & 0.04 & -0.05 & 21 & 0.05 & 0.06 & 19 & 0.06\\
$$[Cu/Fe] & -0.21 & 1 & 0.14 & -0.28 & 1 & 0.12 & -0.21 & 1 & 0.10 & -0.32 & 1 & 0.09 & -0.27 & 1 & 0.13\\
$$[Y\,II/Fe] & -0.05 & 5 & 0.14 & -0.10 & 5 & 0.14 & -0.12 & 3 & 0.15 & -0.03 & 6 & 0.12 & 0.07 & 4 & 0.12\\
$$[Zr\,I/Fe] & 0.56 & 3 & 0.09 & 0.29 & 3 & 0.11 & 0.65 & 2 & 0.08 & 0.31 & 2 & 0.08 & 0.79 & 3 & 0.08\\
$$[Zr\,II/Fe] & 0.45 & 1 & 0.15 & 0.48 & 1 & 0.15 & 0.43 & 1 & 0.15 & 0.45 & 1 & 0.13 & 0.54 & 1 & 0.15\\
$$[Ba\,II/Fe] & 0.20 & 3 & 0.08 & -0.09 & 3 & 0.10 & 0.09 & 3 & 0.10 & -0.06 & 3 & 0.09 & 0.37 & 1 & 0.13 \\
$$[La\,II/Fe] & 0.57 & 2 & 0.20 & 0.39 & 2 & 0.13 & 0.53 & 2 & 0.06 & 0.34 & 2 & 0.04 & 0.59 & 2 & 0.12\\
$$[Ce\,II/Fe] & 0.36 & 2 & 0.27 & 0.10 & 2 & 0.20 & 0.24 & 2 & 0.14 & 0.06 & 3 & 0.19 & -0.10 & 2 & 0.13\\
$$[Nd\,II/Fe] & 0.13 & 5 & 0.13 & 0.24 & 6 & 0.14 & 0.28 & 7 & 0.17 & 0.11 & 8 & 0.12 & ... & ... & ...\\
$$[Eu\,II/Fe] & 0.67 & 1 & 0.15 & 0.38 & 1 & 0.13 & 0.10 &1 &0.14 & 0.26 & 1 & 0.12 & 0.45 & 1 & 0.13\\
\tableline
\end{tabular}
\end{table*}

\begin{figure*}
\centering
\includegraphics[width=1.0\textwidth]{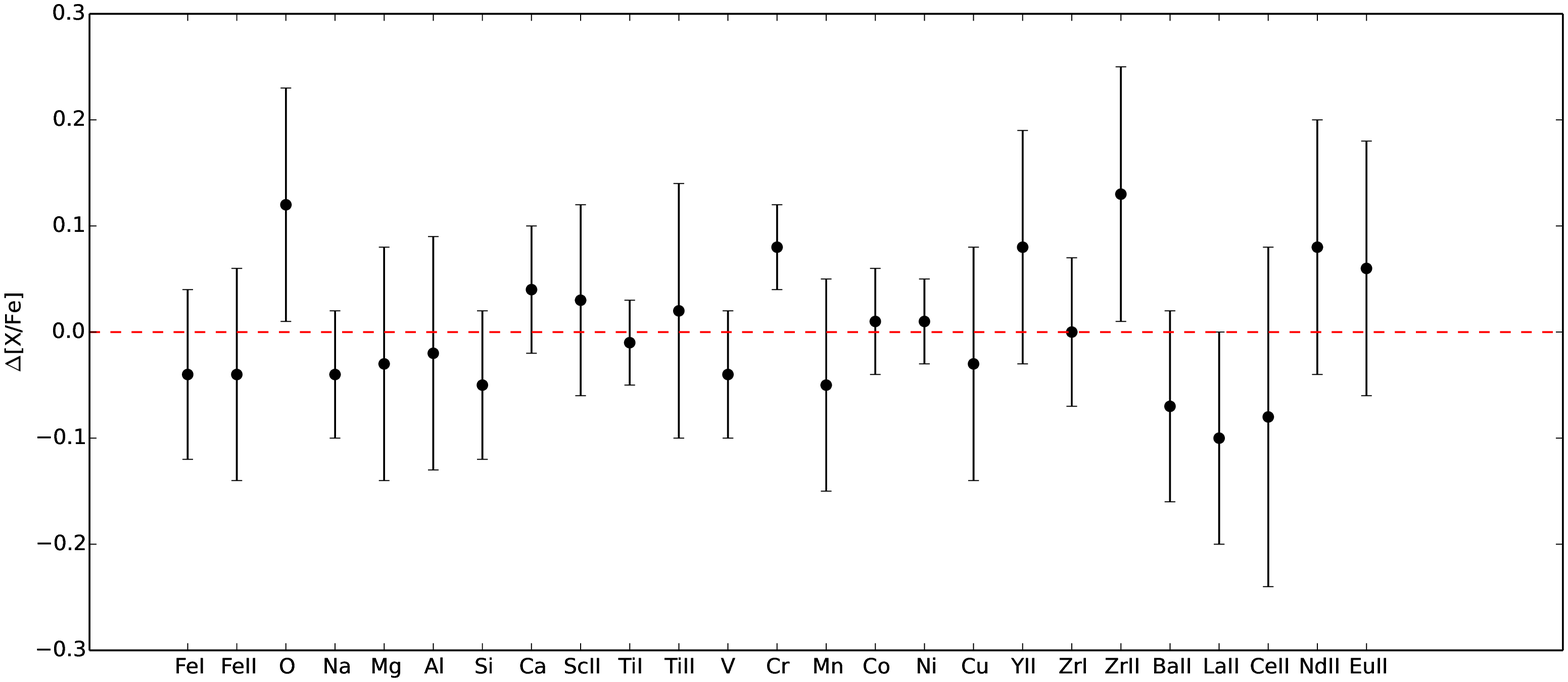}
\caption{Comparison of NGC\,6681\underline{ }12591 abundances derived from Magellan and SALT spectra.  $\Delta$[X/Fe] = [X/Fe]$_{\mathrm{M}}$ - [X/Fe]$_{\mathrm{S}}$. We find good agreement ($\sim0.5\sigma$) between the two sets of abundances for all elements studied.}
\label{fig:Comp12591}
\end{figure*}

Having one star in common between the Magellan and SALT data sets allows us to compare the results of our abundance analysis for each instrument.  We show the results of this comparison in Figure~\ref{fig:Comp12591}, where $\Delta\mathrm{[X/Fe]} = \mathrm{[X/Fe]}_{\mathrm{Magellan}}-\mathrm{[X/Fe]}_{\mathrm{SALT}}$.  The differences between the SALT and Magellan abundances typically agree within 0.5$\sigma$ with the largest differences of 0.13\,dex found in the abundances measure from the fewest lines.  With a typical EW uncertainty of $\sim3.00$\,m\AA\ for S/N$\sim100$, this level of agreement is consistent with the S/N of our EW measurements and implies that there are unlikely to be systematic uncertainties associated with our measurements.

\subsection{Abundance trends in RGBs of NGC\,6681}
\begin{figure*}
\centering
\includegraphics[width=1.0\textwidth]{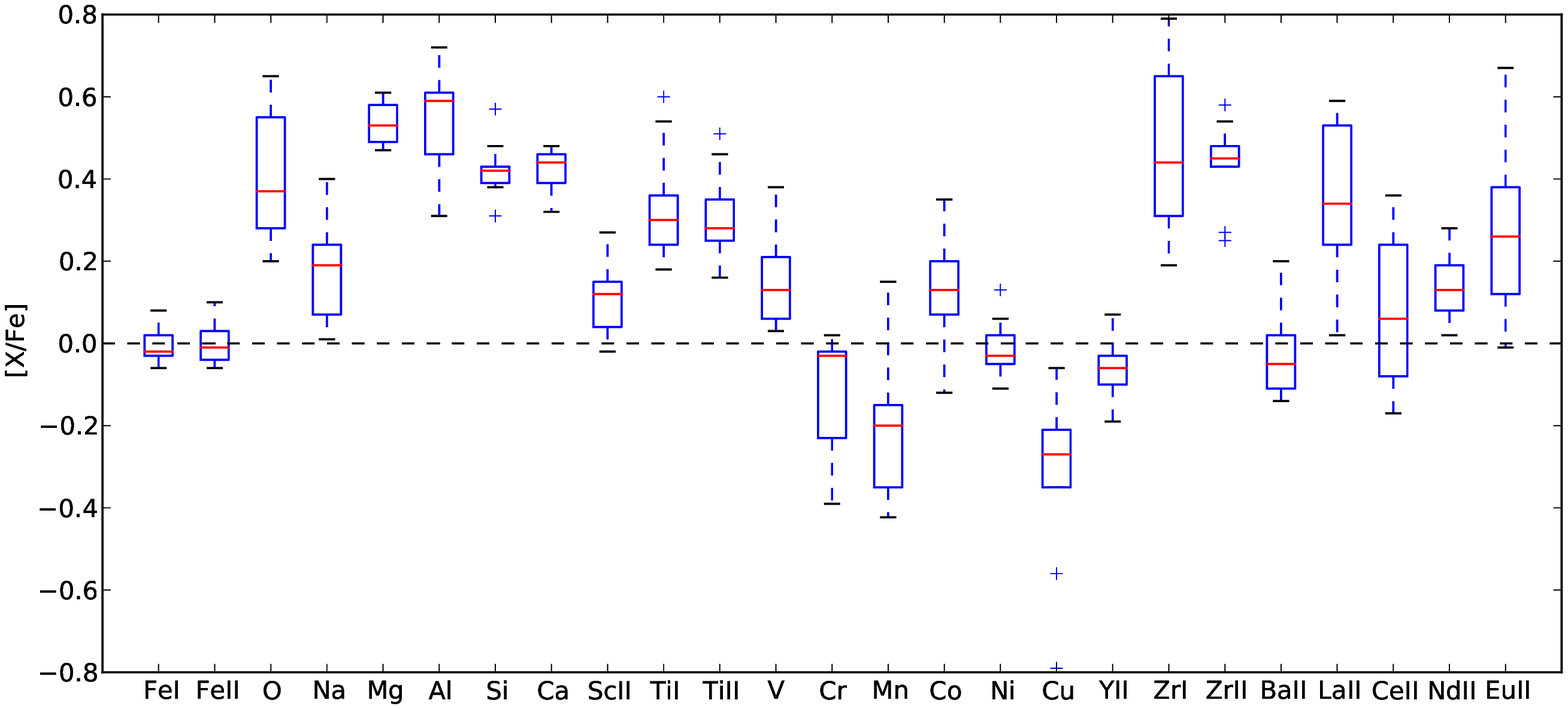}
\caption{Abundance pattern for our sample of RGB stars in NGC\,6681.  The boxes represent the IQR while the horizontal line in each box denotes the median value.  We indicate outliers beyond 1.5 times the second and third quartile as blue crosses.}
\label{fig:whiskers}
\end{figure*}

Before looking at the abundance trends in NGC\,6681 in terms of similarities and differences between individual stars we want to understand the chemical make-up of NGC\,6681 holistically.  The cluster abundance pattern is provided in Figure~\ref{fig:whiskers}.  In this figure, the boxes represent the interquartile range (IQR) (middle 50\%) of our measurements while the horizontal line in each box denotes the median value.  The whiskers indicate either the full abundance range covered by our measurements or 1.5 times the range of the second and third quartile if the full range extends farther.  The few outliers we find are marked with blue crosses.  We present our results in this manner as \citet{Car2006} recommend this as the optimal tool for quantitatively assessing the spread in abundances and performing comparisons between clusters.  We also give these results numerically in Table~\ref{table:IQR} where we provide the median abundance and uncertainty along with the IQR for each element.

\begin{table*}
\centering
\caption{NGC\,6681 abundance statistics. \label{table:IQR}}
\begin{tabular}{l c c c}
\tableline\tableline
Element & Abund & $\sigma_{\mathrm{med}}$ & IQR\\
\tableline
$$[O/Fe] & 0.37 &0.15 & 0.27\\ 
$$[Na/Fe] & 0.19 & 0.12 & 0.17\\
$$[Mg/Fe] & 0.52 & 0.05 & 0.09\\
$$[Al/Fe] & 0.59 & 0.13 & 0.15\\
$$[Si/Fe] & 0.42 & 0.06 & 0.04\\
$$[Ca/Fe] &0.44 & 0.05 & 0.07\\
$$[Sc\,II/Fe] & 0.11 & 0.08 & 0.11\\
$$[Ti\,I/Fe] & 0.29 & 0.13 & 0.12\\
$$[Ti\,II/Fe] &0.28 & 0.10 & 0.10\\
$$[V/Fe] & 0.13 & 0.10 & 0.17\\
$$[Cr/Fe] & -0.03 & 0.14 & 0.21\\
$$[Fe\,I/H] & -1.65 & 0.04 & 0.07\\
$$[Fe\,II/H] & -1.64 & 0.05 & 0.9\\
$$[Mn/Fe] & -0.24 & 0.16 & 0.20\\
$$[Co/Fe] & 0.13 & 0.12 & 0.13\\
$$[Ni/Fe] & -0.04 & 0.07 & 0.07\\
$$[Cu/Fe] &-0.28 & 0.21 & 0.14\\
$$[Y\,II/Fe] & -0.08 & 0.07 & 0.07\\
$$[Zr\,I/Fe] & 0.38 & 0.19 & 0.31\\
$$[Zr\,II/Fe] & 0.45 & 0.10 & 0.24\\
$$[Ba\,II/Fe] & -0.06 & 0.10 & 0.13\\
$$[La\,II/Fe] & 0.34 & 0.17 & 0.36\\
$$[Ce\,II/Fe] & 0.02 & 0.18 & 0.32\\
$$[Nd\,II/Fe] & 0.13 & 0.08 & 0.16\\
$$[Eu\,II/Fe] & 0.19 & 0.19 & 0.26\\
\tableline
\end{tabular}
\end{table*}

\subsection{Light elements}\label{LightTrends}
We are interested in the abundance of light elements in the RGB stars of NGC\,6681 as these elements tend to show abundance patterns indicative not only of GCs, but also of their inherent multiple stellar populations.  In performing the abundance analysis in this study we chose $\alpha$-enhanced model atmospheres as it is expected that for metal-poor GCs we should find $\alpha$-enhancement around 0.2 -- 0.5 dex \citep{GratOrt1986,Barbuy1988,Sneden2004}.  The abundance of most of the $\alpha$-elements is relatively constant, with the exception of O which has a much larger spread, as can be seen in Figure~\ref{fig:whiskers}.  Averaging the abundances of O, Mg, Ca, Si, and Ti for the target stars in this study we find [$\alpha$/Fe] = $0.42\pm0.11$ dex, consistent with what we expect for metal-poor GCs.

The $\alpha$-enhancement we find is not unique to GCs, in fact, the same enhancement is found in metal-poor field stars.  What is unique to GCs are the abundance variations we find among light elements \citep{Cohen1978,Gratton2004,Carretta2009}.  The most noticeable tracer of multiple stellar populations in GCs is the well-known Na-O anti-correlation.  As an element formed through $p$-capture, Na is created in two important ways: in the cores of massive stars as they burn on the MS and at the base of the convective envelope in AGB stars of 3 -- 8$M_\Sun$ \citep{RenzVoli1981,Renz2008}  Therefore, the abundance of Na and its intrinsic variation within GCs can be used to identify multiple stellar populations, with the second generation forming from the Na rich ejecta of the evolved first generation stars \citep{BS1991,Renz2008,MSPrev}.

\begin{figure*}
\centering
\includegraphics[width=1.0\textwidth]{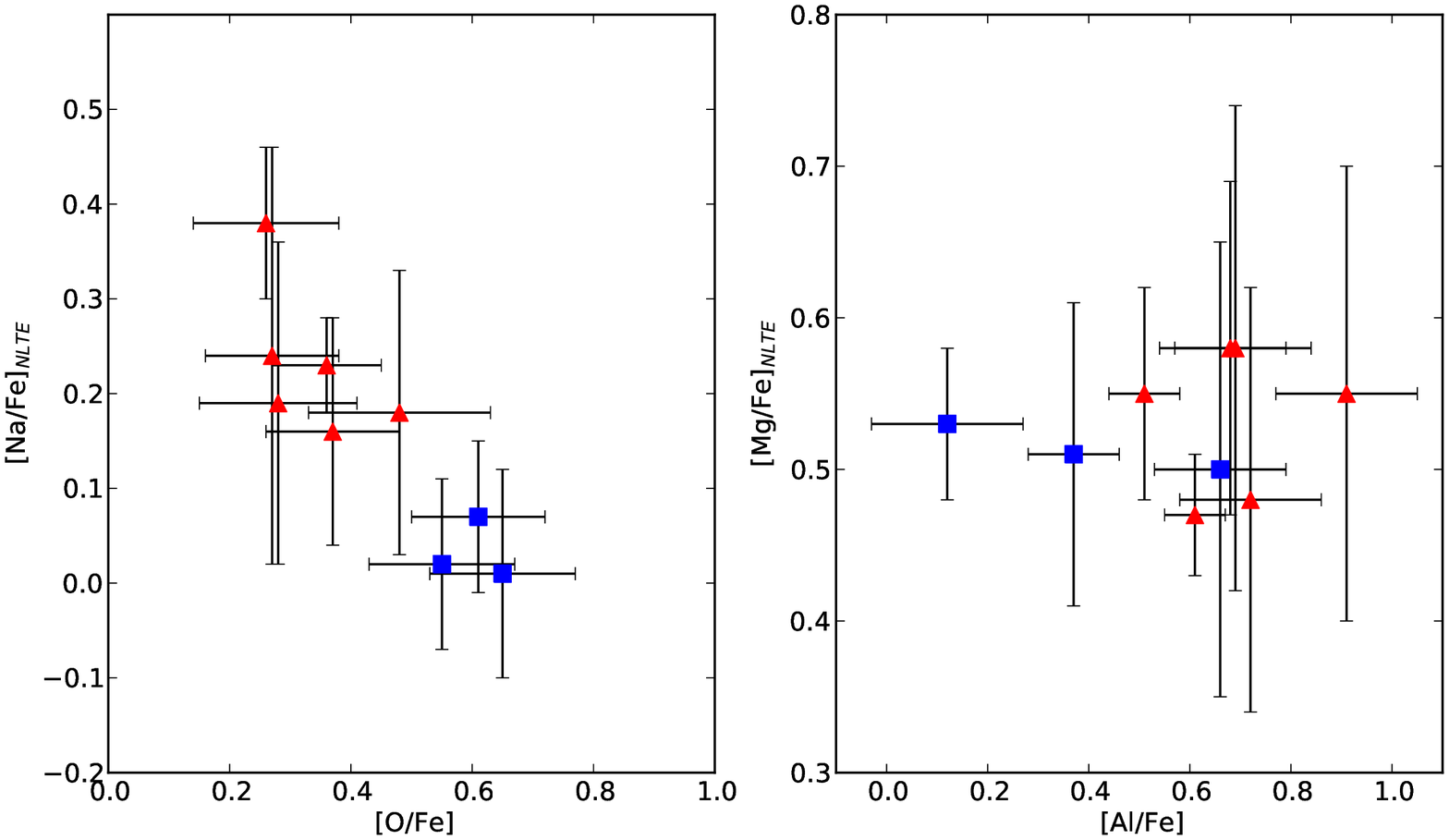}
\caption{Correlations between Na and O (\emph{left}), and Mg and Al (\emph{right}). We adopt a cut of [Na/Fe] = 0.1\,dex to separate the pristine (blue squares) and polluted (red triangles) populations.  A clear Na-O anti-correlation is visible in the left panel, but we do not find evidence of a Mg-Al anti-correlation in NGC\,6681.}
\label{fig:anticor}
\end{figure*}

Spectroscopic evidence suggests that most, if not all, GCs exhibit wide spreads in the Na abundances of their member stars \citep[and references therein]{Carretta2009} and the results of our analysis for NGC\,6681 are consistent with those findings.  We find a 0.39\,dex dispersion in Na and 0.45\,dex in O, both of which are significant compared to the average uncertainty, $\sigma=0.12$\,dex.  We show the Na-O anti-correlation we find in NGC\,6681 in the left panel of Figure~\ref{fig:anticor}.  We use the Kendall rank correlation coefficient, $\tau$, as a statistical measure of the significance in the anti-correlation between Na and O abundances and find $\tau=-0.78$, corresponding to a strong anti-correlation of $p=0.002$ statistical significance.

We are able to identify two populations based on the Na-O anti-correlation found in Figure~\ref{fig:anticor} with separation occurring near [Na/Fe]$\sim0.1$\,dex.  To be certain this is a reasonable estimate we compare the Na abundances of our target stars with those of metal-poor field stars ($-1.8\leq\mathrm{[Fe/H]}\leq-1.4$) from \citet{Bensby2014} and \citet{Johnson2001}.  We do not assume these stars to be exact field counterparts to our cluster stars as the \citet{Bensby2014} sample consists of only dwarf stars and although the \citet{Johnson2001} sample consists of giant stars, they are all more metal-poor than those found in NGC\,6681.  However, the sample as a whole should give a reasonable upper-limit to [Na/Fe] in field stars. We find [Na/Fe]$=-0.09\pm0.06$\,dex in the field sample and take the mean plus three standard deviations as our [Na/Fe] upper-limit.  The population comprised of stars with pristine abundances should have [Na/Fe] similar to that of field stars, whereas the polluted population is any star with [Na/Fe]$\geq0.09$\,dex.  Therefore, our visual estimate of separation in [Na/Fe] at 0.10\,dex is consistent with that predicted by field star abundances.

An anti-correlation between the abundances of Mg and Al has also been shown to exist in some GCs \citep{Grat2001,Carretta2009c,Carretta2010,Meszaros2015}, though it is not as common as the Na-O anti-correlation.  In the right panel of Figure~\ref{fig:anticor} we show the relationship between the abundances of Mg and Al in our RGB stars from NGC\,6681.  We do not find any noticeable trend and in performing a Kendall-$\tau$ test on the data we find a correlation coefficient of $\tau=-0.13$, corresponding to a statistical significance of $p=0.59$.

If the spread in Al we find in our results is intrinsic to the cluster stars and not simply an artifact of a small sample or imperfect treatment of non-LTE effects, then the enhancements we see in Al could be due to proton burning in the MgAl cycle.  In this case, we might expect to see $^{25}$Mg and $^{26}$Mg isotopic enhancements as well.  \citet{Shetrone1996} investigated this and found high values of these Mg isotopes in M13.  More recent studies of NGC\,6752 \citep{Yong2003}, M71 \citep{Yong2006} and $\omega$Cen \citep{DaCosta2013} all show correlations between [Al/Fe] and the abundance of heavy Mg isotopes.  Although MgH lines are visible in the coolest stars in our sample, the lines are too weak and the resolution of our spectra still too low to separate any isotopic components.

\begin{figure*}
\centering
\includegraphics[width=1.0\textwidth]{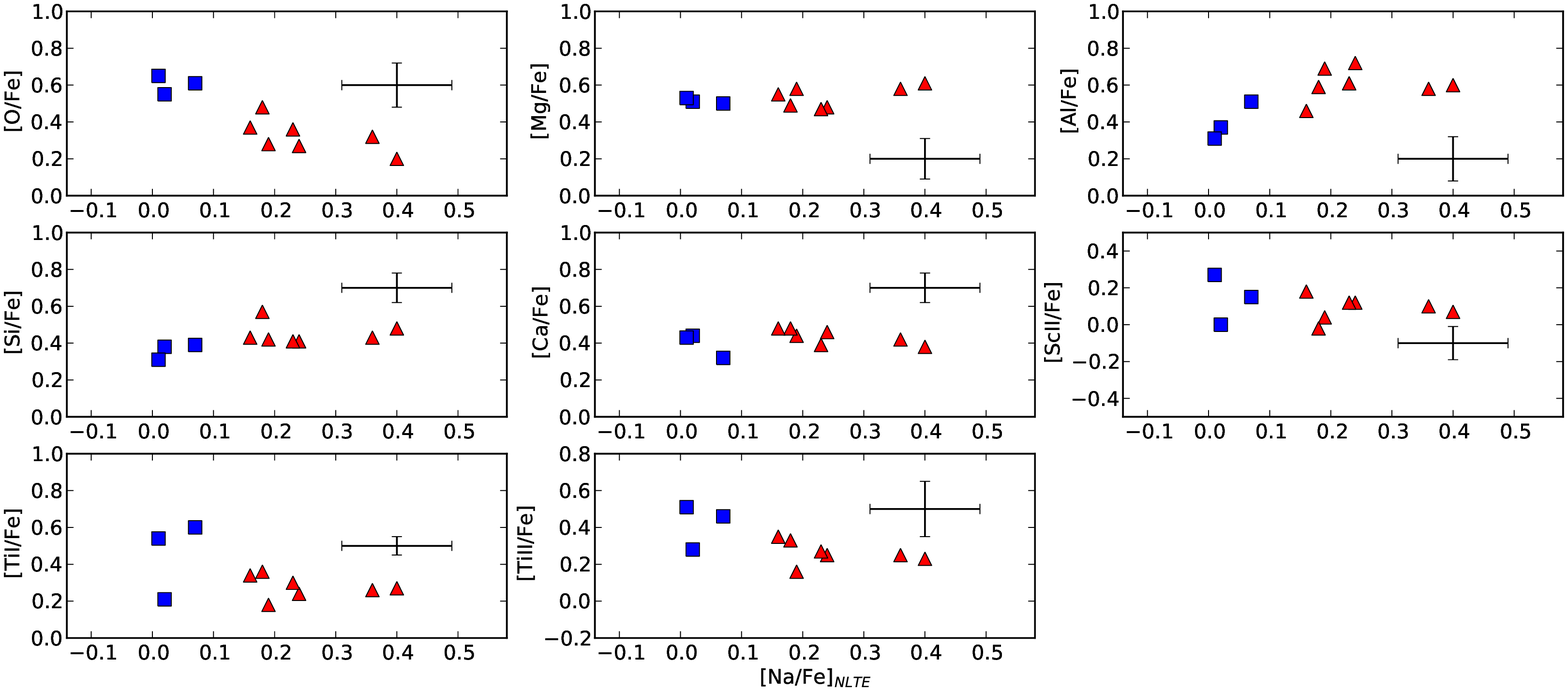}
\caption{[X/Fe] versus [Na/Fe] is shown for the light elements from O to Ti.  The symbols are the same as in Figure~\ref{fig:anticor}.  Along with the individual stellar abundances we denote the representative uncertainties for each element.}
\label{fig:light}
\end{figure*}

Although the Na-O and Mg-Al anti-correlations have been well studied, we are interested to know if any other abundance trends are present in NGC\,6681.  As we define the pristine versus polluted populations based on Na abundances we show in Figure~\ref{fig:light} [X/Fe] versus [Na/Fe] for each light element.  Interestingly, along with O, we find a slight correlation of Na to Al.  However, due to the fact that we have, at most, three Al lines from which to determine the abundance and an incomplete treatment of Al non-LTE processes we cannot be certain this correlation truly represents the relationship between the Na and Al abundances in our stars.

\subsection{Heavy elements}
Through the course of this study we also looked at the abundances of heavier elements, including those in the Fe-peak and $n$-capture groups.  We find the abundances of most of the Fe-peak elements (V, Cr, Mn, Co, and Ni) to be fairly constant among the the RGB stars of NGC\,6681 with an average deviation from the mean of $\sim0.11$\,dex, on the order of the total uncertainty associated with these abundance determinations.  Both Mn and Cr are under-abundant with respect to Fe, whereas Co is slightly overabundant.  These trends are not unusual as GCs such as NGC\,104, NGC\,6752 and NGC\,5272, which cover a range of metallicities, show similar trends among their Fe-peak abundances \citep{Thyg2014,Yong2005,Sneden2004b}.

The scatter in the abundances of many of the $n$-capture elements is noticeably larger than that of the Fe-peak elements, as shown in Figure~\ref{fig:whiskers}.  For the four elements with the largest dispersions (Zr\,I, La\,II, Ce\,II and Eu\,II) the IQR is much larger than the uncertainty of the abundance determinations leading one to believe these are intrinsic spreads and not some artifact of a small sample nor simply a representation of the measurement uncertainty.

\begin{figure*}
\centering
\includegraphics[width=1.0\textwidth]{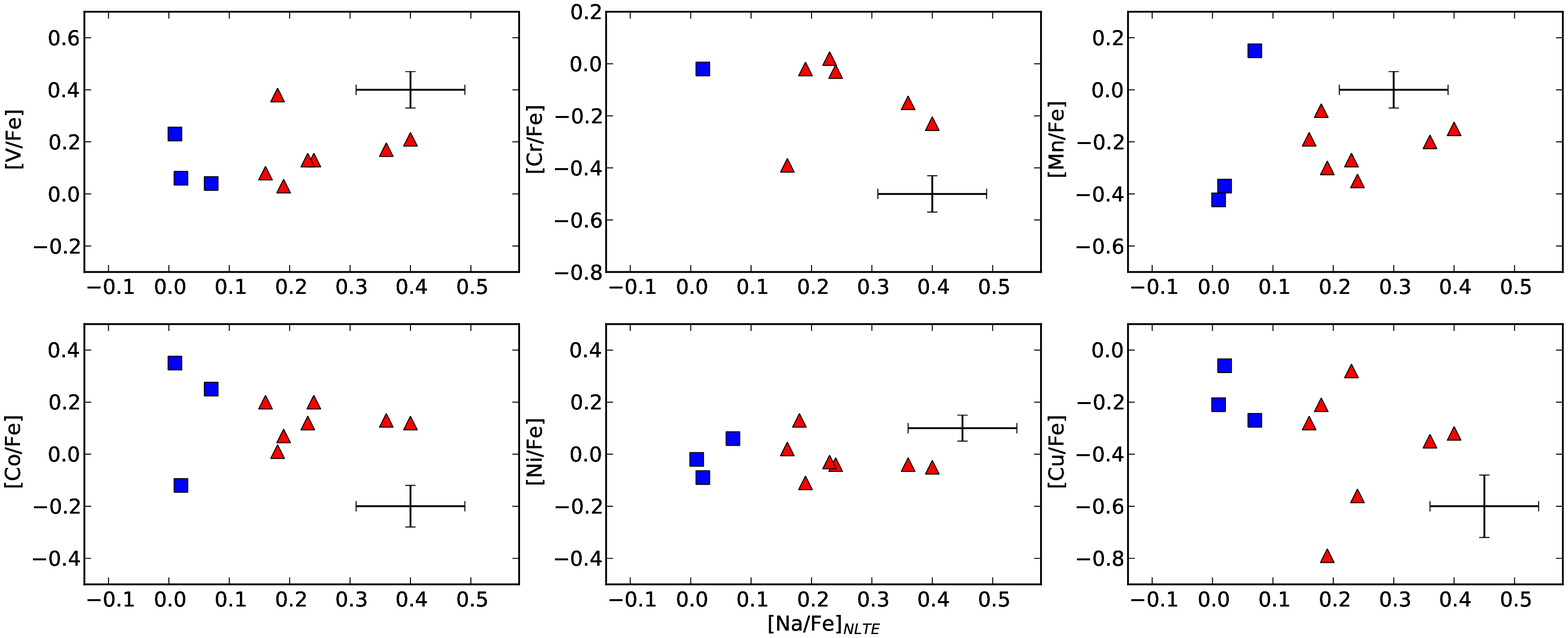}
\caption{[X/Fe] versus [Na/Fe] is shown for the Fe-peak elements from V to Ni and extending to Cu.  The symbols are the same as in Figure~\ref{fig:anticor}.  Along with the individual stellar abundances we denote the representative uncertainties for each element.}
\label{fig:fepeak}
\end{figure*}

\begin{figure*}
\centering
\includegraphics[width=1.0\textwidth]{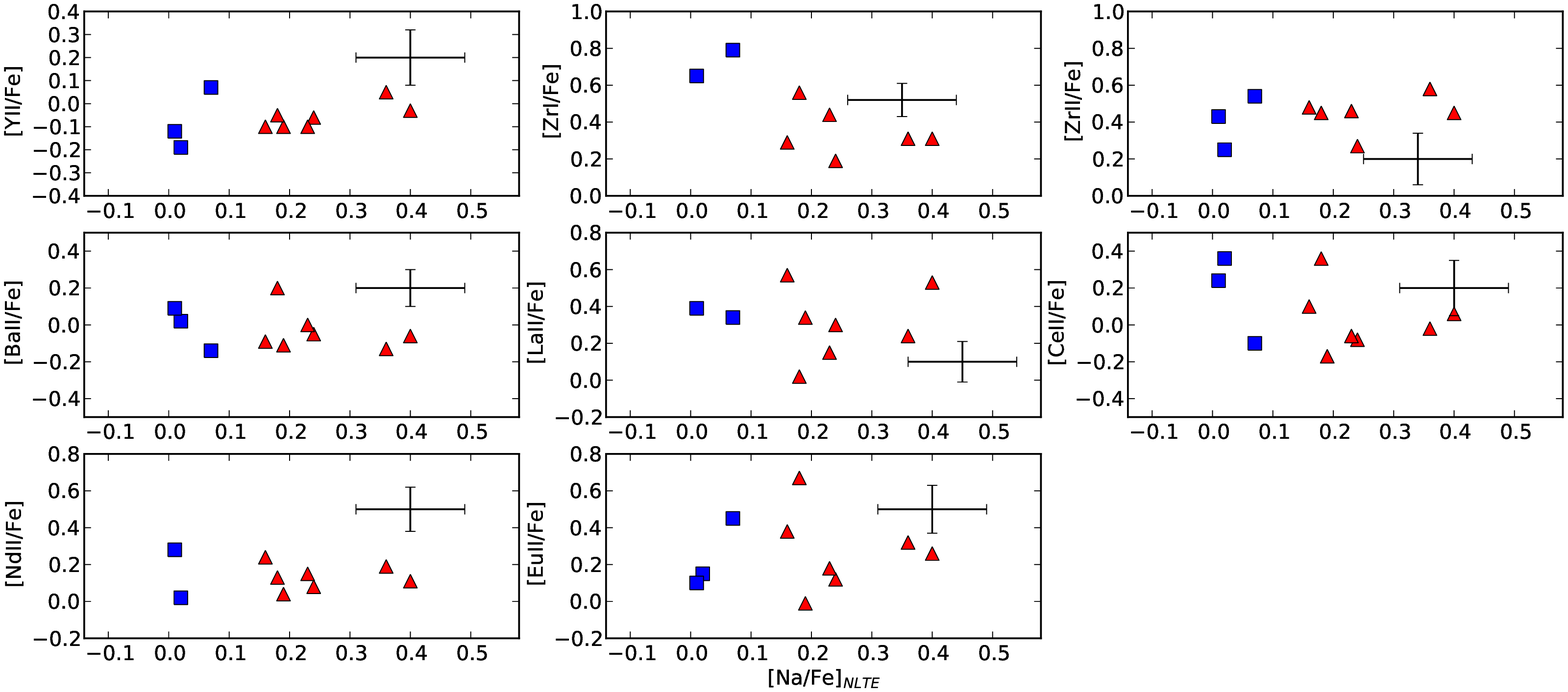}
\caption{[X/Fe] versus [Na/Fe] is shown for the $n$-capture elements from Y\,II to Nd\,II.  The symbols are the same as in Figure~\ref{fig:anticor}.  Along with the individual stellar abundances we denote the representative uncertainties for each element.}
\label{fig:ncapture}
\end{figure*}

As with the light elements, we look for trends between the abundances of Na and the heavier elements.  In Figures~\ref{fig:fepeak} and \ref{fig:ncapture} we do not see any correlations for the Fe-peak elements nor for the majority of the $n$-capture elements, even those with large IQRs.  The exception to this is the $s$-process element Zr\,I, for which the stellar abundances span a range of 0.6 dex and appear to be anti-correlated with those of Na.  In the Sun there are many Zr\,I lines, including the ones used in this study, which have been found to be blended with lines of other elements \citep[and references therein]{Caffau2010}.  The fact that we see a trend in Zr\,I with Na but we do not see a similar trend in Zr\,II with Na leads us to believe that this trend is not intrinsic to the cluster and that Zr\,II is a more reliable abundance indicator.

The Zr\,II abundance is the most enhanced of the $n$-capture elements.  Given this enhancement in Zr, we may also expect enhancement in Rb \citep{McW2013}; however, we do not find evidence of the Rb\,I line at 7800\,\AA.  This is an interesting point as the [Rb/Zr] ratio can be sensitive to the 22Ne($\alpha$,$n$)25Mg reaction in intermediate-mass AGB stars and suggests that future analyses with even deeper spectra of the stars in this cluster could find trace elements of polluter candidates.

We find a significant spread in the abundance of the $r$-process element Eu\,II (larger than that associated with the measurement uncertainties) that does not seem to be correlated with light element abundance dispersions.  Similar abundance variations have been found in M15 \citep{Sobeck2011}, M5, M92 and NGC\,3201 \citep{Roederer2011}.  Specifically, \citet{Roederer2011} find that the $r$-process abundance dispersions are not correlated with light element abundance variations.

To determine the \emph{r}-process enrichment in the cluster we can compare the Eu\,II and Ba\,II abundances of our target stars.  In the solar system, the Eu\,II abundance is dominated by the \emph{r}-process \citep[97\%,][]{Burris2000} while for Ba\,II, the \emph{r}-process contributes much less to the overall abundance but has an 85\% contribution from the \emph{s}-process \citep{Burris2000,Bisterzo2014}.  The [Ba/Eu] abundance ratio has been shown to be a useful diagnostic tool for studying the different neutron capture processes \citep{McWilliam1998} and we find for NGC\,6681 $\mathrm{[Ba/Eu]} = -0.29$\,dex, typical of the MW halo and indicting that both the $s$-process and $r$-process contributed to these elements during halo evolution.  

\begin{figure*}
\centering
\includegraphics[width=0.7\textwidth]{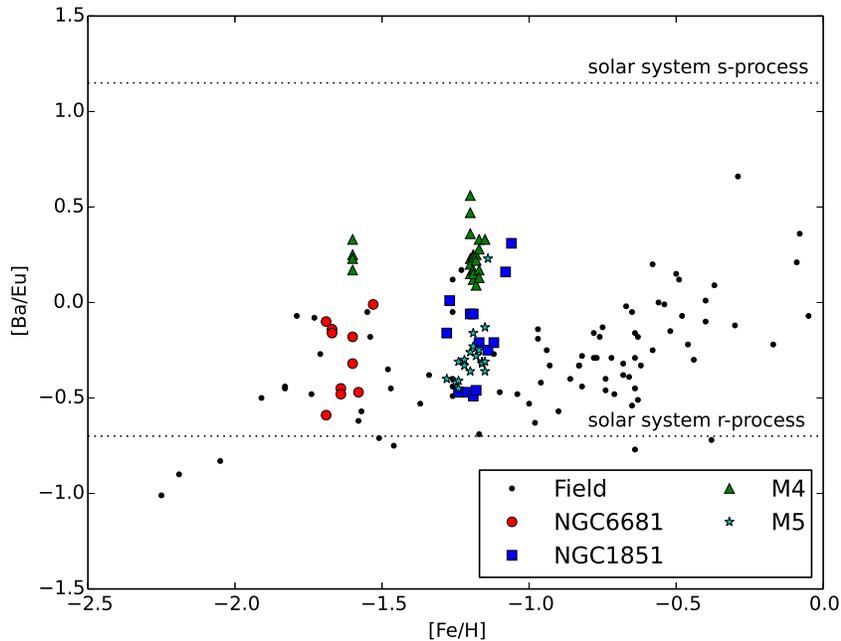}
\caption{[Ba/Eu] versus [Fe/H].  The relationship is shown for field stars (black points) taken from \citet{Fulbright2000}, our NGC\,6681 members (red circles), and the NGC\,1851, M4 and M5 stars in \citet{Sobeck2011}.  The dashed lines depict the solar system \emph{r}-process and \emph{s}-process values based on \citet[and references therein]{Carr2011}.  The mean value for [Ba/Eu] in NGC\,6681 is -0.29 dex.}
\label{fig:BaEu}
\end{figure*}

We show the run of [Ba/Eu] with [Fe/H] for our target stars in Figure~\ref{fig:BaEu}.  The dotted lines show the solar system \emph{s}-process and \emph{r}-process from \citet[and references therein]{Carr2011}.  We also include in this figure field stars (black points) from \citet{Fulbright2000}, and three additional GCs from \citet{Sobeck2011} including NGC\,1851 (blue squares), M4 (green triangles) and M5 (light blue stars).  As expected, we see in this figure that the \emph{s-/r-}process ratio in NGC\,6681 cannot be explained by a purely \emph{r}-process contribution, suggesting again that the \emph{s}-process has played some role in the production of neutron capture elements in NGC\,6681.

\section{Discussion}\label{Discussion}
We obtained high resolution spectroscopic observations of NGC\,6681 to be used in the first detailed abundance analysis of this cluster.  Although we cannot compare our findings to any other results for this cluster other than for [Fe/H], we can compare our abundance results to those of other GCs, specifically those having similar metallicities.  As discussed previously, we find many commonalities between NGC\,6681 and other Milky Way GCs, such as $\alpha$-enhancement, a spread in the Na and O abundances that leads to a distinctive anti-correlation and the identification of two populations of stars, along with an under-abundance of Mn and Cu and overabundance of Co with respect to Fe.  However, unlike many clusters which exhibit signs of an Mg-Al anti-correlation, although we see a spread in Al abundances, this does not translate into a significant trend with Mg abundance in NGC\,6681.

\subsection{Comparison with NGC\,6752}
An interesting cluster for comparison is NGC\,6752 as it has a similar metallicity as reported by \citet{Harris} ([Fe/H]=-1.54\,dex) with some groups finding [Fe/H] as low as -1.61\,dex \citep{Yong2005}.  Additionally, \citet{OMalley2017} find these two clusters to reside in the disk of the Galaxy and report ages of $12.7\pm1.7$ Gyr and $12.6\pm1.7$ Gyr for NGC\,6681 and NGC\,6752, respectively.  Therefore, given their similar [Fe/H] and age, it is possible they may have had similar evolutionary histories.

The light element abundance trends seen in NGC\,6752 as given by \citet{Yong2005} are similar to those we find for NGC\,6681, with larger dispersions seen in the abundances of Na, O, and Al compared to Mg, Si, Ca, Sc\,II, and Ti.  However, the dispersions seen in NGC\,6752 are larger than those seen in NGC\,6681 for Na, O, and Al.  We also find general compatibility with the behavior of heavy element abundances reported by \citet{Yong2005}, specifically the considerable under-abundance of both Mn and Cu as compared to Fe, slight under-abundance of [Y\,II/Fe] and overabundance in the remaining heavy elements.

\citet{Yong2005} perform a similar comparison of \emph{n}-capture \emph{s}-process and \emph{r}-process mechanisms and find [Ba/Eu] = -0.37 dex which falls between the pure \emph{r}-process and solar mix of \emph{s}+\emph{r} and is slightly more \emph{r}-process dominated than the result we find for NGC\,6681.  Additionally, we find larger abundances of the \emph{s}-process elements Zr and La in NGC\,6681.  As these elements are produced in the interiors of AGB stars through $n$-capture $s$-process, these elements can be used as tracers of this mechanism in understanding the polluters associated with multiple stellar populations in GCs.  The greater enhancement of Zr and La abundances we see in NGC\,6681 lends support to a greater $s$-process contribution in this cluster.

\subsection{Photometric evidence of multiple stellar populations}
In the present study we have been able to identify two distinct populations of stars in NGC\,6681, a pristine sample which is poor in Na but rich in O, and a polluted sample which is Na-rich and O-poor.  The abundance difference between the pristine and polluted samples suggest that the stars comprising the polluted population were formed not from the original cluster material, but from material that has been processed by an earlier generation of stars.  For many GCs, there is also strong photometric evidence for multiple stellar populations which manifests itself in the form of split evolutionary sequences in a CMD if the right combination of photometric filters is used.

It was shown in \citet{Milone2012} and \citet{Monelli2013} that the color index $c_{U,B,I} = (U-B)-(B-I)$ is an effective tool for disentangle multiple stellar populations in GCs as it incorporates both the (U-B) color which is sensitive to abundance variations among light elements \citep{Marino2008} and the (B-I) color which, due to its long wavelength baseline, highlights temperature differences and is capable of separating populations with different He abundances \citep{Piotto2007}.  \citet{Monelli2013} found a small but significant spread in the RGB of NGC\,6681, giving us reason to believe this GC is host to multiple stellar populations.  

\begin{figure*}
\centering
\includegraphics[width=0.7\textwidth]{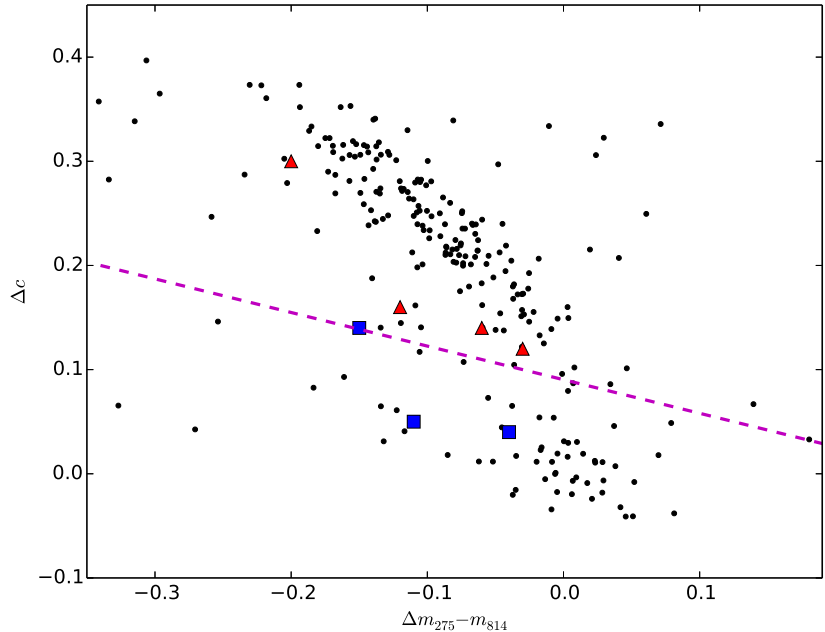}
\caption{We show the chromosome map depicting $\Delta c_{m275,m336,m438}$ versus $\Delta m_{275},m_{814}$ as prescribed in \citet{Milone2017} for NGC\,6681.  Our Na-poor (pristine) target stars are marked in blue while the Na-rich (enhanced) stars are marked in red.  The dotted line separates first generation, Na-poor, stars second generation, Na-rich, stars.}
\label{fig:chrome}
\end{figure*}

Additionally, subsequent papers \citep{Milone2015a,Milone2015b,Milone2017} have used the combination of the pseudo-color, $c_{m275,m336,m438}$ and the $m_{275}-m_{814}$ color to maximize the separation between stellar populations and, in doing so, have constructed ``chromosome maps" for 57 GCs including NGC\,6681.  We follow the prescription for constructing a chromosome map present in \citet{Milone2017}, using \emph{HST} optical and UV data from \citet{Sara2007} and \citet{PiottoCMD}.  We present our results in Figure~\ref{fig:chrome} and identify the location of our seven target stars with UV data.

A clear separation can be seen in Figure~\ref{fig:chrome} where first generation, Na-poor, stars are located below the dotted line while second generation, Na-rich stars, are found above the line.  \citet{Milone2017} also find that only 23\% of the stars in NGC\,6681 are first generation stars and our results would agree with this finding.  We find that our spectroscopic population identification for our target stars agrees well with the photometric separation evident in this figured.  The one curious result is that of NGC\,6681\underline{ }30380 which falls on the separating line and we are therefore not able to confirm the spectroscopic results using this photometric test.

\subsection{Possible Pollution Mechanisms}
There are several scenarios in the literature for polluter candidates that attempt to explain the existence of multiple stellar populations in GCs including fast rotating massive stars (FRMS), massive interacting binaries (MIB), super massive stars (SMS) and asymptotic giant branch (AGB) stars \citep{Renzini2015}.  Recently, the AGB scenario has found more favor and the results of our analysis provide supporting evidence against the other three candidates.  

\citet{DH2014} propose SMS as polluter candidates as these stars are expected to be fully convective, thereby expelling a homogeneous stellar wind that would progressively enrich the interstellar medium (ISM) with processed material (i.e. CNO-cycle material and $p$-capture products).  However, the mass range required to produce Na enrichment is too low to produce O depletions of the scale seen in observations, including those seen here in NGC\,6681.  Another issue with this scenario is that it does not eliminate the possibility for Type Ia supernova (SN Ia) to occur and enrich the ISM with heavy elements prior to the formation of second generation stars.  We do not find a significant dispersion in the [Fe/H] values of the cluster members that would support first generation SN Ia having polluted the material from which second generation stars formed.

\citet{Krause2013} admit a different difficulty in their scenario of FRMS which also impacts the MIB scenario of \citet{deMink2009}, which is that they will produce a continuous abundance spread and not discrete populations.  We have seen in this study, both spectroscopically and photometrically, two distinct populations of stars in NGC\,6681 with unique chemical abundance patterns.  With high resolution spectroscopy, \citet{Carretta2014} has also been able to define discrete populations in NGC\,2808, adding confidence to our findings in NGC\,6681.  As with SMS, FRMS and MIB do not eliminate the possibility of Type Ia SN in the cluster which leads us to eliminate these as possible formation scenarios for multiple stellar populations based on our abundance determinations.  Additionally, at least for FRMS, it is expected one would find an $\alpha$-enhancement in the second generation of stars compared to the first \citep{Conroy2011} and we do not find observational evidence of that to be true.  

AGB stars are expected to be good polluter candidates as they undergo the process of hot bottom burning in which the bottom of the convective envelope reaches high enough temperatures for the $p$-capture nuclear processing required to create variations in light element abundances in subsequent generations.  Multiple star-formation events can occur as the AGB stars eject their outer layers, forming second generation stars before the gas reservoir in the cluster is removed by the first Type Ia SNs; therefore, the [Fe/H] content will remain constant between first and second generation stars which agrees with our [Fe/H] measurements.  

\citet{Ventura2016} performed the first investigation of the plausibility of the AGB scenario using Mg and Al abundances in GCs.  The authors note that the Mg-Al anti-correlation is an even better constraint on polluter candidates than the Na-O anti-correlation as the MgAl cycle occurs at higher temperatures and the abundances in RGB stars are not affected by mixing.  The findings suggest that Mg depletion will only occur in low mass, metal-poor AGB stars.  We do not see any depletion in the Mg abundances in NGC\,6681 when studying their spectra at visible wavelengths; however, we do not use this as evidence against the AGB scenario.  The Mg-Al anti-correlation is not a ubiquitous trait seen in all GCs and therefore a our confidence in a pollution model should not rest on this relation alone.  Also, it is expected that near-IR spectroscopy would be better for this analysis as Mg, Al, and Fe are less effected by non-LTE in this wavelength region \citep{GH2015}.

Although the AGB scenario has gained popularity in recent years, it is not without shortcomings.  The most obvious problem with AGB stars as polluter candidates is that, taken alone, they produce a correlation between Na and O abundances \citep{Dantona2011}, not the anti-correlation seen in observations.  Many models of AGB polluters have tried to address these problems \citep[e.g.][]{Dercole2010,Dercole2016,Conroy2011}; however, a solution has yet to be found that reproduces all observations.  The remaining issues with the AGB scenario, and very likely the other polluter scenarios, such as the mass budget problem and the necessity of a pristine gas reservoir for dilution, are beyond the scope of this study as they cannot be explained directly by the results of our abundance analysis.

\section{Summary}\label{Summary}
We obtained the first high resolution spectroscopic observations of the Galactic GC NGC\,6681 in order to derive detailed abundances for 23 elements in nine RGB stars in this cluster.  We confirm the membership of these stars based on radial velocity measurements of $214.5\pm3.7$\,km\,s$^{-1}$, which is consistent with previous studies, \citep{BB1994,Rosenberg2000,Francis2014} and are therefore able to draw conclusions about the existence and properties of multiple stellar populations based on the abundance trends of these cluster members.

Our abundance analysis focuses on not only Fe and $\alpha$-elements, which tend to be the most well-studied, but also other light elements, Fe-peak elements, and $n$-capture elements.  We use a spectroscopic approach to deriving stellar atmosphere parameters such that we obtain both abundance and ionization equilibrium for Fe.  Although the majority of our abundances have been determined under the assumption of LTE, some of the elements studied require a non-LTE approach and we have made corrections where appropriate.  The only element that deserves such non-LTE treatment that was not corrected is Al as adequate models are not currently available in the parameter space covered by our target stars.

We find an mean cluster metallicity of $\mathrm{[Fe/H]}=-1.63\pm0.07$\,dex which is in agreement with previous studies \citep{Carretta2009b,Saviane2012} and an $\alpha$-enhancement of [$\alpha$/Fe] = $0.42\pm0.11$ dex, consistent with what is expected for metal-poor GCs.  Additionally, we confirm the existence of a Na-O anti-correlation in NGC\,6681 and, in doing so, are able to identify two stellar populations with the polluted population having $\mathrm{[Na/Fe]}>0.1$\,dex and the pristine population having $\mathrm{[Na/Fe]}<0.1$\,dex.  We do not find evidence for a Mg-Al anti-correlation, which has been seen in some clusters.  Previous studies have found interesting correlations between [Al/Fe] and the heavy isotope abundances of Mg; however, the resolution of our target spectra is too low to resolve any isotopic components.  Future studies of higher resolution might investigate this correlation.

In our investigation of heavy elements in NGC\,6681, we find no correlation between the Fe-peak nor the $n$-capture element abundances and the light elements that exhibit abundance variations, with the exception of Zr\,I which we do not believe to be intrinsic to the cluster.  We find a [Ba/Eu] ratio that is greater than purely \emph{r}-process, suggesting a combination of both $r$-process and $s$-process mechanisms.

Finally, with the combination of the spectroscopic abundances found in this study and photometric data available in multiple \emph{HST} filters \citep{Sara2007,PiottoCMD,Soto2017}, we are able to form a nearly complete picture of the nature of the multiple stellar populations in NGC\,6681.  We can make a direct connection between the discrete populations we see photometrically on the RGB and the abundances in different stellar populations.  The particular mechanism responsible for the abundance variations we see is still an area of active research and these results should help put constraints on the possible polluter candidates.

\acknowledgements
The authors thank the anonymous referee for giving constructive scientific comments and suggestions to improve the manuscript.  This material is based upon work supported by the National Science Foundation Graduate Research Fellowship under Grant No. DGE-1313911.  Any opinion, findings, and conclusions or recommendations expressed in this material are those of the authors(s) and do not necessarily reflect the views of the National Science Foundation.  A.K. acknowledges support from the National Research Foundation (NRF) of South Africa and  by the Russian Science Foundation (project no. 14-50-00043).  Some of the observations reported in this paper were obtained with the Southern African Large Telescope (SALT) under program codes 2015-1-SCI-070 and 2016-1-SCI-002.  This paper also includes some data gathered with the 6.5-m Magellan Clay Telescope located at Las Campanas Observatory, Chile.

\end{document}